\documentclass[fleqn,usenatbib]{mnras}

\usepackage{newtxtext,newtxmath}
\usepackage{longtable}
\usepackage[T1]{fontenc}

\DeclareRobustCommand{\VAN}[3]{#2}
\let\VANthebibliography\thebibliography
\def\thebibliography{\DeclareRobustCommand{\VAN}[3]{##3}\VANthebibliography}

\usepackage{graphicx}	

\title[HORuS transmission spectroscopy and revised planetary parameters of KELT-7 b]{HORuS transmission spectroscopy and revised planetary parameters of KELT-7 b}

\author[H. M. Tabernero et al.]{
H. M. Tabernero$^{1}$\thanks{E-mail: htabernero@cab.inta-csic.es},
M. R. Zapatero Osorio$^{1}$,
C. Allende Prieto$^{2,3}$,
E.  Gonz{\'a}lez-{\'A}lvarez$^{1}$,
J. Sanz-Forcada$^{4}$
\newauthor
A. L{\'o}pez-Gallifa$^{5,1}$,
D. Montes$^{5}$,
C. del Burgo$^{6}$,
J. I. Gonz{\'a}lez Hern{\'a}ndez$^{2,3}$,
and R. Rebolo$^{2,3}$
\\
$^{1}$ Centro de Astrobiolog{\'i}a (CSIC-INTA), Carretera de Ajalvir km 4, E-28850 Torrej{\'o}n de Ardoz, Madrid, Spain\\
$^{2}$ Instituto de Astrof{\'i}sica Canarias, C/ V\'{i}a L\'{a}ctea s/n, E-38205 La Laguna, Tenerife, Spain\\
$^{3}$ Universidad de La Laguna, Dpto. Astrof\'{i}sica, E-38206 La Laguna, Tenerife, Spain\\
$^{4}$ Centro de Astrobiolog{\'i}a (CSIC-INTA), ESAC Campus, Camino bajo del Castillo s/n, E-28692 Villanueva de la Ca\~nada, Madrid, Spain\\
$^{5}$ Departamento de F{\'i}sica de la Tierra y Astrof{\'i}sica \& IPARCOS-UCM (Instituto de F\'{i}sica de Part\'{i}culas y del Cosmos de la UCM), \\Facultad de Ciencias F{\'i}sicas, Universidad Complutense de Madrid, E-28040 Madrid, Spain\\
$^{6}$ Instituto Nacional de Astrof\'{i}sica, \'{O}ptica y Electr\'{o}nica, Luis Enrique Erro 1, Sta. Ma. Tonantzintla, Puebla, Mexico\\
}

\date{Accepted XXX. Received YYY; in original form ZZZ}

\pubyear{2021}

\begin{document}
\label{firstpage}
\pagerange{\pageref{firstpage}--\pageref{lastpage}}
\maketitle
\begin{abstract}

We report on the high-resolution spectroscopic observations of two planetary transits of the hot Jupiter KELT-7b ($M_{\rm p}$~$=$~1.28~$\pm$~0.17~M$_{\rm Jup}$, $T_{\rm eq}$~$=$~2028~K) observed with the High Optical Resolution Spectrograph (HORuS) mounted on the 10.4-m Gran Telescopio Canarias (GTC). A new set of stellar parameters are obtained for the rapidly rotating parent star from the analysis of the spectra. Using the newly derived stellar mass and radius, and the planetary transit data of the Transiting Exoplanet Survey Satellite ({\sl TESS}) together with the HORuS velocities and the photometric and spectroscopic data available in the literature, we update and improve the ephemeris of KELT-7b. Our results indicate that KELT-7 has an angle $\lambda$~$=$~$-$10.55~$\pm$~0.27~deg between the sky projections of the star's spin axis and the planet's orbital axis.  By combining this angle and our newly derived stellar rotation period of 1.38 $\pm$ 0.05 d, we obtained a 3D obliquity $\psi$ = 12.4 $\pm$ 11.7 deg (or 167.6 deg), thus reinforcing that KELT-7 is a well-aligned planetary system. We search for the presence of H$\alpha$, \ion{Li}{i}, \ion{Na}{i}, \ion{Mg}{i}, and \ion{Ca}{ii} features in the transmission spectrum of KELT-7b but we are only able to determine upper limits of 0.08--1.4~\%~on their presence after accounting for the contribution of the stellar variability to the extracted planetary spectrum. We also discuss the impact of stellar variability in the planetary data. Our results reinforce the importance of monitoring the parent star when performing high-resolution transmission spectroscopy of the planetary atmosphere in the presence of stellar activity.

\end{abstract}

\begin{keywords}
planets and satellites: atmospheres -- planets and satellites: individual: KELT-7b -- methods: observational -- techniques: spectroscopic -- stars: activity
\end{keywords}



\section{Introduction}
\label{sec:intro}

Planets orbiting main-sequence stars are ubiquitous as demonstrated by the more than 4\,000  exoplanets discovered to the present day\footnote{\url{https://exoplanetarchive.ipac.caltech.edu}}. Among them, transiting exoplanets are of great interest because they allow us to investigate the structure and bulk chemical composition of their upper atmospheres \citep{2013A&ARv..21...63T}. In particular, highly-irradiated gaseous planets are key targets for atmospheric characterisation due to their proximity to the parent host-stars, transit depth and duration. \citet{char02} reported the first detection of an exoplanet atmosphere by means of space-based transmission spectroscopy. However, the recent works by \citet{cas20, 2021A&A...647A..26C} have raised some doubts over this detection showing that  at high spectral resolution it is mostly explained by the Rossiter-McLaughlin (R-M) effect. Despite this setback, \citet{char02} opened a new era in the characterisation of exoplanetary atmospheres that has flourished thanks to high spectral resolution spectrographs such as CARMENES \citep{quir16}, ESPRESSO \citep{pepe20}  HARPS \citep{may03}, HARPS-N \citep{cos12}, HIRES \citep{vog94}, and MAROON-X \citep{sei16}. The first successful attempts to explore the presence of neutral sodium in the atmospheres of HD~209458~b and HD~189733~b with ground-based facilities were reported by \citet{sne08} and \citet{red08}. Over the next decade, many ultra-hot Jupiters (UHJs), defined as giant gaseous planets with equilibrium temperatures above $\sim$2\,200 K \citep[see][]{par18}, were explored via the transmission spectroscopy technique using ground- and space-based instruments. \ion{Fe}{i,ii} and \ion{Ti}{i,ii} features have been detected in the atmosphere of the  UHJ KELT-9~b \citep{hoe18,hoe19}. In addition, iron has also been detected in other UHJs such as MASCARA-2b \citep{cas19}, WASP-33b \citep{nug20,cont21}, WASP-76b \citep{ehr20,tab21b,kes21}, and WASP-121b \citep{gib20,hoe20,bor21}. Other chemical species (He, Li, Na, Mg, Ca, K, V, Cr, CO, CH$_4$, and H$_2$O) have been reported in the atmospheres of tens of highly irradiated gaseous planets  \citep[e.g.][]{barm15,wyt15,she17,chen18,nor18,par18,all18,all19,alo19,sei19,sing19,yan19,hoe20,tab21b,bor21b}.

Spectroscopy is a powerful tool used to explore the presence of atomic and molecular  species in the atmospheres of transiting exoplanets.  The High Optical Resolution
Spectrograph\footnote{\url{http://www.gtc.iac.es/instruments/hors/horus.php}}  \citep[HORuS, see][]{horus} at the 10.4-m Gran Telescopio Canarias (GTC) offers an excellent opportunity to probe the transmission spectra of transiting planets, given its moderately high resolving power of $R$~$\approx$~25\,000 and the large collecting area of the telescope, all of which are key for achieving a high signal-to-noise ratio (S/N) during the brief time window of planetary transits. 

The Jupiter-sized planet KELT-7b orbits an F2V-type star (HD~33643, $V$~$=$~8.54~mag) with an orbital period of 2.7~d  and a mass of 1.3~M$_{\rm Jup}$ \citep{bier15}. KELT-7b is an inflated hot Jupiter with an equilibrium temperature of 2\,050~K \citep[see][]{bier15,zhou16} that moves around a  modestly metal-rich ([Fe/H]~$=$~0.14~$\pm$~0.08~dex) fast-rotating star \citep{bier15}. This planet reunites all required properties for performing transmission spectroscopy. In this regard, \citet{plu20} presented the first study of KELT-7b's atmosphere using near-infrared data obtained with the {\sl Hubble Space Telescope}. They detected the presence of H$_2$0 and H$^-$ absorption in the planetary atmosphere.  However, to the best of our knowledge, the optical transmission spectrum of KELT-7b has not been explored so far. This fact together with the high equilibrium temperature of this planet with the  suspected high metallicity of the parent star (that may indicate a high abundance of metals in the planetary atmosphere), make it an excellent target for atmospheric characterisation at optical wavelengths.  In this work, we use high S/N HORuS spectra of KETL-7 to study the Rossiter-McLaughlin (R-M) effect, the Doppler Shadow, and the transmission spectrum of the planet. We also revise the planetary ephemeris and parameters by taking advantage of multiple {\sl TESS} light curves alongside the publicly available radial velocity (RV) measurements. We also explore the activity level of the star using X-ray data, photometry and spectroscopy.
 
This manuscript is organised into the following sections: the observational data analysed in this work are presented in Sect.~\ref{sec:obs}, the characterisation of the KELT-7 system can be found in Sect.~\ref{sec:par}, the Rossiter-McLaughlin (R-M) analysis can be found in Sect.~\ref{sec:RM}, the transmission spectrum is analysed in Sect.~\ref{sec:trans}, and the conclusions and final remarks appear in Sect.~\ref{sec:concfinal}.

\section{Observations}
\label{sec:obs}

\subsection{HORuS spectra}

Two transits of KELT-7b were observed on the nights of 2019 November 17 (first transit, hereafter T1) and 2020 January 30 (second transit, T2) with HORuS at the 10.4-m GTC of the Spanish Observatorio del Roque de los Muchachos on the island of La Palma. These spectroscopic observations were gathered under the framework of the Spanish open time allocation commission under proposal 126-GTC122/19B (PI H. M. Tabernero).  HORuS was built with components from the Utrecht Echelle Spectrograph\footnote{\url{https://www.ing.iac.es/PR/wht_info/whtues.html}} (UES), which was in operation at the 4.2-m William Herschel Telescope (WHT) between 1992 and 2001. It collects light at the Nasmyth focal plane, shared with OSIRIS\footnote{Optical System for Imaging and low-Intermediate-Resolution Integrated Spectroscopy; \url{http://www.gtc.iac.es/instruments/osiris/}}, using a 3x3 image slicer ($2.1\times2.1$ arcsec) into optical fibres that form a pseudo-slit at the spectrograph entrance, using microlenses on both ends. The light is dispersed with a 79 gr~mm$^{-1}$ {\it echelle} grating and cross-dispersed with three prisms, providing almost continuous wavelength coverage between approximately 3\,800 and 6\,900 \AA{}. The HORuS image slicer makes the instrument tolerant of mediocre seeing conditions ($<$20\% loss of light from point sources for a seeing of FWHM~1.5 arcsec). In T1 and T2, we collected a total of 76 and 60 consecutive HORuS spectra, respectively,  with an individual exposure time of 300~s.  In addition to the science images, we acquired Th-Ar exposures intended for wavelength calibration with an exposure time of 3.75~s. The average S/N of each individual exposure is about 180 per pixel at $\sim$~5\,500~\AA{}.  HORuS delivers optical spectra in the wavelength interval 3\,770--6\,910~\AA{} with a nominal resolving power of $R \approx 25\,000$. We note that the wavelength coverage is not continuous and there are some gaps between the {\it echelle} orders towards the red.  We used the planetary ephemeris of \citet{bier15} to schedule the observations: in principle, our objective was to observe the system continuously from one hour before through one hour after the transit. However, this strategy was hampered by the uncertainties of the ephemeris; we needed to update the planetary ephemeris (see below). The log of the spectroscopic observations is given in Table~\ref{tab:obs}.

The spectroscopic data were reduced using the {\tt chain}\footnote{\url{https://github.com/callendeprieto/chain}}, a custom-made data reduction pipeline for HORuS written in IDL. Raw data were bias subtracted; the pipeline traced and optimally extracted the apertures corresponding to the various spectral orders. Wavelength calibration involved fitting third or second-order polynomials to one to two dozen thorium lines in each {\it echelle} order, identified in an atlas made with the Tull spectrograph \citep{cap01}, and interpolating linearly the wavelength solutions between the Th-Ar exposures preceding and following each stellar spectrum.

\subsection{{\sl TESS} and KELT photometry}

To improve the ephemeris and planetary parameters of KELT-7b, we gathered the photometric light-curves of KELT-7 obtained by the Transiting Exoplanet Survey Satellite ({\sl TESS}) mission. KELT-7 (TIC~367366318) was observed by \textit{TESS} in 2 min short-cadence integrations in Sector 19 during the {\it TESS} primary mission,  and Sectors 43, 44, and 45 during the extended mission. The \textit{TESS} data are publicly available as target pixel files (TPFs) and calibrated light curves (LCs), being provided in FITS format. These files contain a primary Header Data Unit (HDU) with more metadata stored in the header. The first extension HDU contains more metadata in the header, and stores arrays of data in a binary FITS table, which includes the timestamps, Simple Aperture Photometry (SAP) fluxes, and Pre-search Data Conditioned Simple Aperture Photometry (PDCSAP) fluxes. The SAP flux is computed by summing the fluxes in the calibrated pixels within the \textit{TESS} optimal photometric aperture, while the PDCSAP flux corresponds to the SAP flux values nominally corrected for instrumental variations. Thus, the PDCSAP flux values are the best available estimate of the intrinsic variability of the target and are optimised for \textit{TESS} transit searches. All {\sl TESS} data were processed by the Science Processing Operations Center pipeline \citep[SPOC,][]{2016SPIE.9913E..3EJ}.

The {\sl TESS} photometry was joint together with the light curves of the Kilodegree Extremely Little Telescope\footnote{\url{https://keltsurvey.org}} \citep[KELT, ][]{kelt} survey published by \citet{bier15}. The {\sl TESS} data were downloaded from the Mikulski Archive for Space Telescopes\footnote{\url{https://archive.stsci.edu/}} (MAST) which is a NASA founded project, whereas the KELT light-curves were collected from the NASA exoplanet archive\footnote{\url{https://exoplanetarchive.ipac.caltech.edu/}}. 

\subsection{TRES radial velocities}

As a complement to the {\sl TESS} and KELT photometric data, we gathered the publicly available radial velocities (RVs) determined by \citet{bier15} using the Tillinghast Reflector Echelle Spectrograph \citep[TRES, see][]{tres}. These RVs include one transit of KELT-7b and we divided them into two categories.  The first category corresponds to the RVs outside the planetary transit which are employed in a combined photometric and radial velocity analysis (Section~\ref{sec:plapar}). They have an average uncertainty of 87.6~m~s$^{-1}$, an $rms$ of 119.4~m~s$^{-1}$ and cover a time span of 752~d. Finally, the second category contains in-transit RVs which are employed to study the classical Rossiter-McLaughlin (R-M) effect alongside the HORuS spectra (see Section~\ref{sec:classrm}.)\\

\subsection{{\sl XMM-Newton} data}

We acquired {\sl XMM-Newton} observations through DDT program ID 85338 (PI J. Sanz-Forcada) on 2019 Aug 29, to better evaluate the stellar activity of KELT-7, and possible effects of photoevaporation in the planet atmosphere (see Sect.~\ref{sec:concfinal}). The 8.5 ks observation resulted in an effective exposure time of 5.5, 8.0, and 8.1 ks for the PN, MOS1, and MOS2 EPIC cameras on board {\sl XMM-Newton}, respectively. The target is detected with a combined S/N~$=$~11.7. A coronal model spectra was simultaneously fit to the data of all three cameras ($\log T$ (K) $= 6.32^{+0.28}_{-0.27}$, $6.80^{+0.22}_{-0.17}$, $\log EM$ (cm$^{-3}$) $= 50.58^{+0.37}_{-0.52}$, $50.71^{+0.21}_{-0.85}$), assuming the photospheric abundance of [Fe/H]~$=$~0.24~dex (see Table~\ref{par_planet}). The unabsorbed X-ray luminosity of KELT-7 is $L_{\rm X}= (4.1\pm0.4)\times 10^{28}$~erg\,s$^{-1}$ as measured in the ``standard'' 0.12--2.48 keV spectral range (5--100~\AA), assuming an interstellar medium (ISM) absorption of $N_{\rm H} = 10^{20}$~cm$^{-3}$ in the direction of KELT-7. We modelled the stellar emission in the range 100--920~\AA, following \citet{sf11}, resulting in $L_{\rm EUV}=6.2\times 10^{29}$~erg~s$^{-1}$. The expected luminosity in the range 100--504~\AA{} is $5.2\times 10^{28}$~erg~s$^{-1}$. The Optical Monitor (OM) telescope onboard XMM-Newton registered a flux of  $F_{\rm UVW2}=1.7368\pm0.0048 \times 10^{-13}$~erg\,s$^{-1}$\,cm$^{-2}$\,\AA$^{-1}$\ in the $UVW2$ filter centred at 2120~\AA\ (bandwidth 500~\AA) for KELT-7.

\section{The KELT-7 system}
\label{sec:par}

\subsection{Stellar parameters}
\label{sec:stepar}

\begin{figure*}                                      
\centering                       
    \centerline{\includegraphics[width=0.5\textwidth]{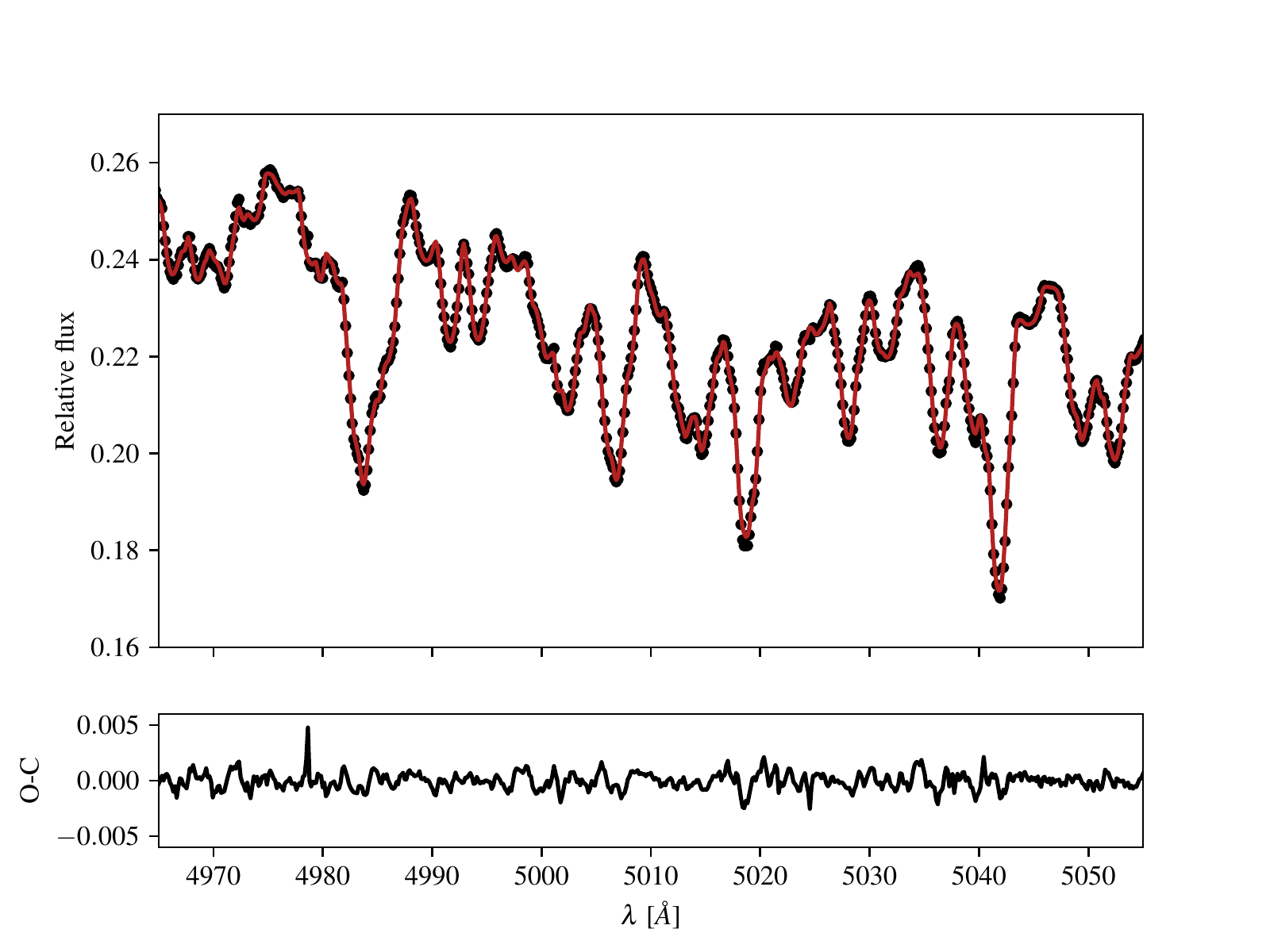}
     \includegraphics[width=0.5\textwidth]{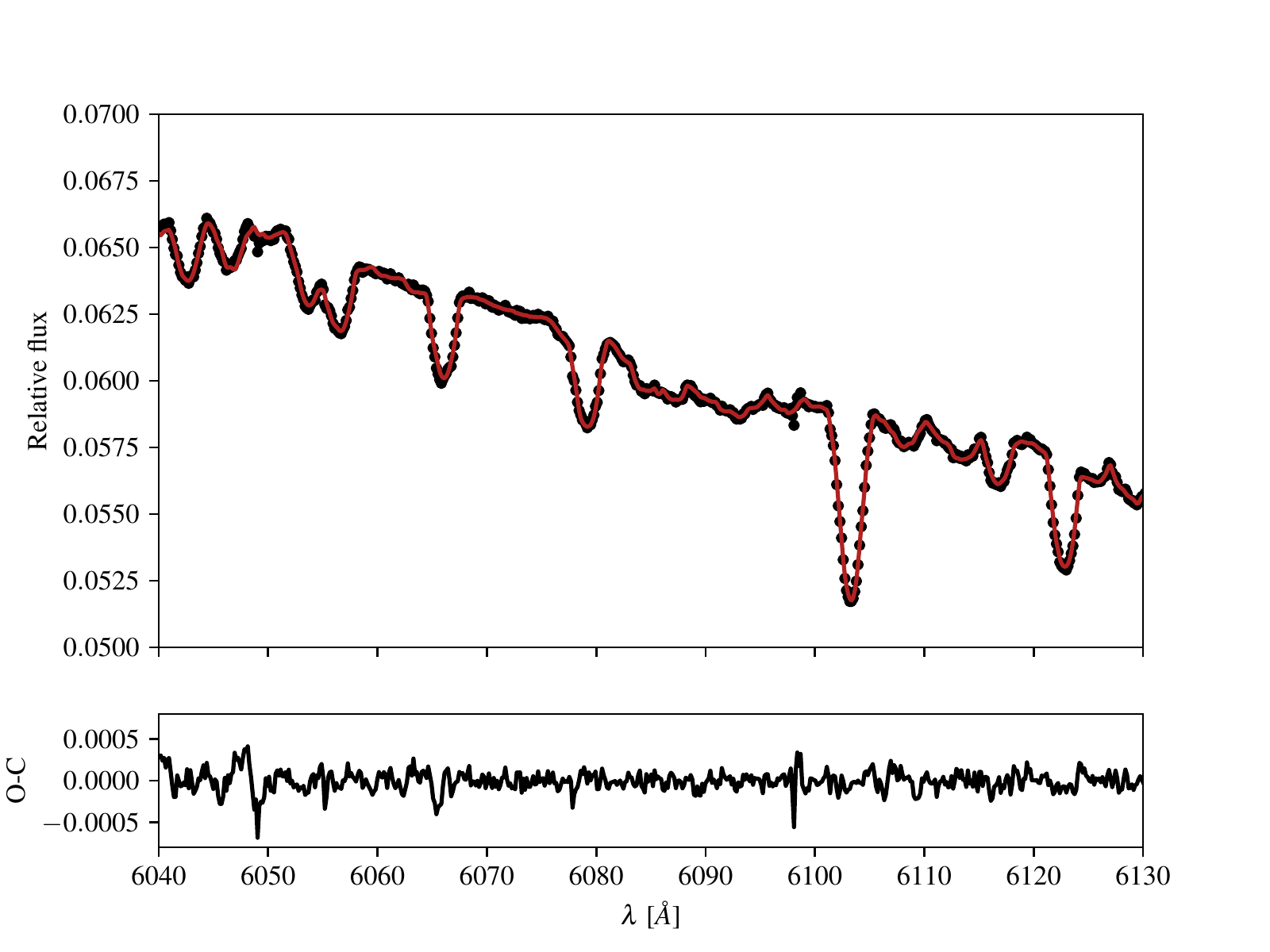}}
    \caption{Two different spectral regions of the HORuS combined spectrum of KELT-7 (black dots) are compared to the best-fit theoretical spectrum (red line) with the stellar parameters given in Table~\ref{par_planet}. The model was broadened by using the spectroscopic rotational velocity of the star. The residuals of the fit (O-C) are given in the bottom panels. The main atomic features are due to \ion{Ti}{i}, \ion{Ni}{i}, and \ion{Fe}{i}. Wavelengths are in the air system.}
    \label{best_fit_syn}
\end{figure*}

We computed the stellar atmospheric parameters of KELT-7, namely: effective temperature ($T_{\rm eff}$), surface gravity ($\log{g}$), metallicity ([Fe/H]), and the projected rotational velocity ($\varv \sin{i}$) using the spectral synthesis method \citep[see, e.g.][]{nisgus18}. To that aim, we adopted a grid of synthetic spectra to reproduce the HORuS combined spectrum in the range 3\,800--6\,800~\AA{}. The grid of synthetic spectra was computed using the BT-Settl model atmospheres \citep{all12}, the radiative transfer code {\tt Turbospectrum}\footnote{\url{https://github.com/bertrandplez/Turbospectrum2019}} \citep{turbospectrum}, and a VALD3\footnote{\url{http://vald.astro.uu.se/}} atomic line list. The grid spans over the following parameter ranges: 6\,000~$<$~$T_{\rm eff}$~$<$~8\,000~K, 3~$<$~$\log{g}$~$<$~5~dex, $-1~<$~[Fe/H]~$<$~0.5~dex, and the microturbulence ($\xi$) of each model is fixed to the values given by the calibration of \citet{dutraferr}, which depends only on $T_{\rm eff}$ and $\log{g}$. In order to derive the spectroscopic rotational velocity $\varv \sin{i}$, we modelled the observed line broadening with two components following the approach described in \citet{lumba}. The first component accounts for the stellar $\varv \sin{i}$ that we modelled thanks to the rotation profile described in \citet{gra08}. The second component accounts for the instrumental line spread function (LSF) of HORuS and it was described by a Gaussian function with a FWHM corresponding to the instrumental resolving power. Then, we employed the {\sc SteParSyn} code\footnote{\url{https://github.com/hmtabernero/SteParSyn}} \citep{tab22} and the BT-Settl model spectra to retrieve the following stellar atmospheric parameters and their associated uncertainties: $T_{\rm eff}$~$=$~6\,699~$\pm$~24~K, $\log{g}$~$=$~4.15~$\pm$~0.09~dex, [Fe/H]~$=$~0.24~$\pm$~0.02~dex, and $\varv \sin{i}$~$=$~71.4~$\pm$~0.2~km~s$^{-1}$. We display the best fitting models in two representative spectral regions in Fig.~\ref{best_fit_syn}. \\ 

Using these stellar parameters we calculated the stellar age, mass ($M_{*}$) and radius ($R_{*}$) of KELT-7 with the PARAM web interface\footnote{\url{http://stev.oapd.inaf.it/cgi-bin/param}} \citep{sil06}. We used $T_{\rm eff}$, [Fe/H], the {\sl Gaia} EDR3 Parallax \citep{EDR3}, the visual magnitude ($V$), and the PARSEC stellar evolutionary tracks and isochrones \citep{bre12}. We obtained $M_{*}$~$=$~1.517~$\pm$~0.022~M$_{\odot}$, $R_{*}$~$=$~1.712~$\pm$~0.037~R$_{\odot}$, and age~$=$~1.2~$\pm$~0.7~Gyr. In addition to these values, PARAM delivers an independent determination of the stellar surface gravity of $\log{g}$~$=$~4.13~$\pm$~0.01~dex that is fully-consistent with the spectroscopic value within the quoted uncertainties. 

We also employed the Virtual Observatory SED Analyser\footnote{\url{http://svo2.cab.inta-csic.es/theory/vosa/}}  \citep[VOSA, see][]{vosa} to produce the photometric spectral energy distribution (SED) of KELT-7 from the $U$- through the $W4$-band using publicly available data: the $UBV$ photometry from \citet{UBV}, the $G$, $G_{BP}$, $G_{RP}$ magnitudes from {\sl Gaia} DR3 \citep{EDR3}, the $JHK$ magnitudes from 2MASS \citep{2MASS}, and the $W1-W4$ magnitudes from {\sl WISE} \citep{WISE}. We then fitted the observed SED using BT-Settl and ATLAS9 models \citep{all12,atlas09} and found the following stellar parameters: $T_{\rm eff}$~$=$~6700~$\pm$~100~K, $\log{g}$~$=$~4.0\,$\pm$\,0.5~dex, and [Fe/H]~$=$~0.2~$\pm$~0.2~dex that are fully consistent with those derived using {\sc SteParSyn}. Additionally, we measured the bolometric luminosity of the star by integrating over the best SED model to be $L_{\rm bol}$~$=$~5.359\,$\pm$\,0.077~$L_{\odot}$. All adopted stellar parameters are summarised in Table~\ref{par_planet} and will be used in our reanalysis of the KELT-7 planetary system. 

In order to validate our stellar parameters, we scrutinised the literature and found two other parameter determinations for KELT-7. The first determination corresponds to \citet{bier15} and they reported the following measurements: $T_{\rm eff}$~$=$~6\,789$^{+49}_{-50}$~K, $\log{g}$~$=$~4.15~$\pm$~0.02~dex, [Fe/H]~$=$~0.14~$\pm$~0.08~dex, and $\varv \sin{i}$~$=$~65.0$^{+5.9}_{-6.0}$~km~s$^{-1}$. The second determination was performed by  \citet{zhou16} and they provided: $T_{\rm eff}$~$=$~6\,513$^{+49}_{-53}$~K, $\log{g}$~$=$~4.13~$\pm$~0.03~dex, and $\varv \sin{i}$~$=$~69.3~$\pm$~0.2~km~s$^{-1}$. Our $\log{g}$ is compatible within error bars with the two available determinations. Only \citet{bier15} provided a value for [Fe/H] that in turn is compatible at the 1~$\sigma$ level with our values. Our $T_{\rm eff}$ is bracketed by these two literature values and is compatible at the 2~$\sigma$ level with the value reported by \citet{bier15}, whereas the $T_{\rm eff}$ calculated by \citet{zhou16} deviates from our determinations by almost 200~K. The $\varv \sin{i}$ calculated using {\sc SteParSyn} is compatible with that of \citet{bier15} and the value presented in \citet{zhou16} only differs by 2.1~km~s$^{-1}$. Regarding the mass and radius for KELT-7, \citet{bier15} gave $M_{*}$~$=$~1.535$_{-0.054}^{+0.066}$~M$_{\odot}$, $R_{*}$~$=$~1.732$^{+0.045}_{-0.043}$~R$_{\odot}$, and age~$=$~1.3~$\pm$~0.2~Gyr, all of them being compatible with the PARAM determinations within the error bars. Interestingly, both determinations provide a value of $R_{*}$ that is unusually large for an F2 star in the main sequence, suggesting that KELT-7 might be a slightly evolved star. This is still compatible with the age of 1.2~$\pm$~0.7~Gyr we derived with the PARAM web interface. In fact,  an F2~V star should have $\log{g}$~$=$~4.3, whereas an F2~III star has a gravity of 3.7~dex according to the values tabulated in \citet{cox00}. KELT-7 has a $\log{g}$ of 4.13~dex that is constrained by these two values. Therefore, we infer that KELT-7 has already left the main sequence and is entering the sub-giant phase.

\begin{table}
\caption{Adopted orbital and physical parameters for the KELT-7 system}            
\centering
\label{par_planet}      
\begin{tabular}{lcc}   
\hline
Parameter &  Value & Reference \\    
\hline                       
& Stellar properties & \\
 \hline
   $\alpha$ (J2000) &  05:13:10.94    &  {\it Gaia} EDR3 \\
   $\delta$ (J2000) & +33:19:04.61    &  {\it Gaia} EDR3 \\
   SpT & F2~V & \citet{1993yCat.3135....0C} \\
   $V$ & 8.54~mag & \citet{bier15} \\
   $T_{\rm eff}$    & 6699~$\pm$~24~K & This work \\  
   $\log{g}$  &  4.15~$\pm$~0.09~dex &   This work \\
   $\lbrack$Fe/H$\rbrack$  & 0.24~$\pm$~0.02~dex   &  This work \\ 
   $\xi$  &  1.29~$\pm$~0.10~km~s$^{-1}$ & This work \\
   $\varv \sin{i}$ & 71.4~$\pm$~0.2~km~s$^{-1}$ & This work \\
   $M_{*}$   &  1.517~$\pm$~0.022~M$_{\odot}$  &  This work \\
   $R_{*}$  &  1.712~$\pm$~0.037~R$_{\odot}$ & This work\\
   $\rho_{*}$     & 0.425~$\pm$~0.035 g~cm$^{-3}$ & This work \\
   $L_{\rm X}$ & (4.1~$\pm$~0.4) $\times$ 10$^{28}$ erg s$^{-1}$ & This work \\
   $L_{\rm bol}$ & 5.359~$\pm$~0.077~L$_{\odot}$  & This work\\
   $\log L_{\rm X}/L_{\rm bol}$ & $-$5.70~$\pm$~0.05 & This work\\
   Age       & 1.2~$\pm$~0.7~Gyr & This work\\
   $\pi$     & 7.38~$\pm$~0.02~mas& {\it Gaia} EDR3 \\
   $d$ & 135.5~$\pm$~0.4 pc & {\it Gaia} EDR3 \\
   RV & 40.7~$\pm$~0.8 km~s$^{-1}$ &  {\it Gaia} EDR3 \\
   $\mu_{\alpha}$~$\cos{\delta}$ & 10.477~$\pm$~0.024~mas~yr$^{-1}$& {\it Gaia} EDR3\\
   $\mu_{\delta}$ &   	$-$49.620~$\pm$~0.018~mas~yr$^{-1}$ & {\it Gaia} EDR3 \\
   $d$      & 135.5~$\pm$~0.4~pc & {\it Gaia} EDR3 \\
   $U$ & $-$43.4~$\pm$~0.8~km~s$^{-1}$ & This work\\
   $V$ & $-$24.3~$\pm$~0.1~km~s$^{-1}$ & This work\\
   $W$ & $-$15.6~$\pm$~0.1~km~s$^{-1}$ & This work\\
   $P_{\rm {rot}}$ & 1.38 $\pm$ 0.05 d & This work  \\
   \hline 
& Planet properties &\\
\hline
  $P$  & 2.73476550~$\pm$~0.00000033~d & This work \\ 
  $a$  &  0.0434~$\pm$~0.0012~au & This work \\
  $a/R_{*}$ &  5.452~$\pm$0.028  & This work \\
  $i$  & 83.51~$\pm$~0.09~deg & This work \\
  $T_{14}$ &   3.48~$\pm$~0.04~h & This work \\   
  $\delta$ & 	 8022~$\pm$~22~ppm & This work\\
  $R_p/R_{*}$    & 0.08957~$\pm$~0.00012 & This work \\
   $M_{\rm p}$  &  1.28~$\pm$~0.17~M$_{\rm Jup}$  &  This work\\
   $R_{\rm p}$  &  1.496~$\pm$~0.035~R$_{\rm Jup}$  &  This work  \\
   $K$    &  140$\pm$~17~m~s$^{-1}$ & This work  \\
   $T_0$     &   2458835.661885$\pm$~0.000073~BJD & This work \\
   $\rho_p$ & 0.48~$\pm$~0.10~g~cm$^{-3}$ & This work \\
   $e$       & 0  &   adopted\\
   $\omega$  & 90~deg &  adopted \\
   $T_{\rm eq}$ & 2028~$\pm$~17~K &  This work\\
\hline
\end{tabular}
\end{table}

\subsection{Radial velocities}
\label{radvel}

We determined the radial velocity (RV) of each HORuS spectrum by means of the cross-correlation method implemented in the {\tt iSpec} code \citep{bla14} and using a reference mask. We calculated the cross-correlation functions (CCFs) as a function of observing time with a set of line masks corresponding to an F0~V star, which has a spectral type similar to our target. Regions affected by strong telluric lines were masked out. Then, we sampled the CCFs from $-$300 to 300 km~s$^{-1}$ with a step of 0.5~km~s$^{-1}$. KELT-7 is a fast-rotating star with a $\varv \sin{i}$ of $71.4$~km~s$^{-1}$, which implies quite broadened CCFs that are not well reproduced by a Gaussian function: the Gaussian fit to the CCFs thus delivers RVs with large uncertainties \citep[e.g.,][]{bor19}. Consequently, we modelled the CCFs with a Gray rotation function \citep{gra08} convolved with a Gaussian kernel corresponding to the HORuS resolving power. The fitting was performed with the Levenberg-Marquardt algorithm (LMA) implemented in the {\tt curve-fit} routine of the {\tt SciPy} Python package \citep{scipy}. This method is robust and it delivers both the RVs and its corresponding uncertainties. The typical uncertainty of the HORuS RVs is at the level of 200 m\,s$^{-1}$.  The list containing the individual HORuS RVs is given in Table~\ref{tab:obs}.\\

The final systemic RV of KELT-7 was taken from the {\sl Gaia} EDR3 (Table~\ref{par_planet}). Using the systemic RV, stellar coordinates, and proper motion listed in Table~\ref{par_planet}, we computed the Galactocentric space velocities as in \citet{mon18}, which includes the equations and methodology described in \citet{uvw87}. The Bayesian analysis tool of \citet{2018ApJ...856...23G} indicates that KELT-7 has a very high probability (99.9\,\%) of being a  field star and not belonging to any of the 27 nearby young stellar associations considered by the algorithm.  For its space velocities, KELT-7 has a likely kinematical age~$\ge$~0.8~Gyr, consistent with the age determined in  Section~\ref{sec:stepar}.

\subsection{Stellar activity}
\label{sec:activity}

\begin{figure}                                      
\centering                       
    \includegraphics[width=0.55\textwidth]{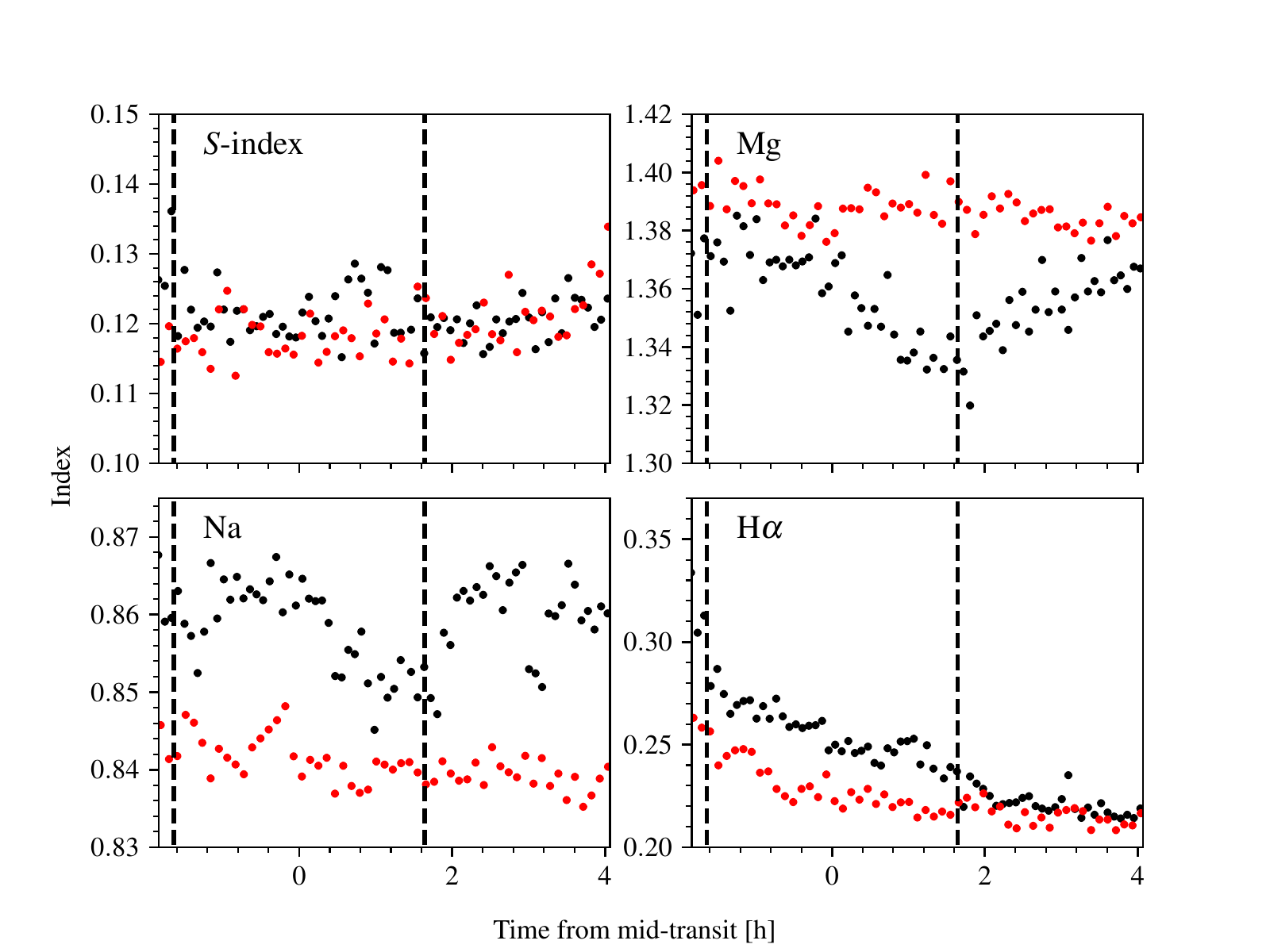}
    \caption{
    Activity indices for the \ion{Ca}{ii} H and K (S-index), \ion{Mg}{i}b triplet, \ion{Na}{i} D, and H$\alpha$ atomic lines shown as a function of time for the two observed transits (T1 and T2 are plotted as black and red dots, respectively). The ingress and egress times are marked with a vertical dashed line whereas the mid-transit time for both T1 and T2 has been set to 0~h.}
    
    \label{activity}
\end{figure}

Before proceeding to the extraction of the planetary transmission spectrum, we inspected the activity state of the parent star in HORuS T1 and T2 separately. More specifically, we intended to determine the evolution of the stellar activity during the planetary transits because stellar activity variations may have an important impact in the analysis and retrieval of the planetary signal. We calculated spectral indices centred at \ion{Ca}{ii}~H\&K (the $S$-index), H$\alpha$, \ion{Na}{i} D, and \ion{Mg}{i} b atomic features. The first three indices were measured by using the definitions given in \citet{gomdasil11,gomdasil21}. For the Mg index, however, we used a custom definition because HORuS covers two lines (5167.32 and 5172.68~\AA) out of the three of the triplet, while the third line is too close to the edge of the detector. Our Mg index is thus calculated as the quotient of the flux in the centre of the two covered Mg lines (in a wavelength interval of 0.5~\AA{} width to map the core of the lines) divided by the average flux in two adjacent continuum regions with a width of 10~\AA{} centred at around 5\,110 and 5\,205~\AA{}, effectively following the philosophy of the indices definition given in \citet{gomdasil11,gomdasil21}. We represent the time series of all four spectral indices in Fig.~\ref{activity}. All of these optical indices explore the stellar activity at different atmospheric heights: H$\alpha$ and \ion{Ca}{ii}~H\&K trace the middle and lower chromosphere while \ion{Na}{i} D, and \ion{Mg}{i} b typically trace the upper photosphere and lower chromosphere.

We found that the $S$-index had the same values for the two transits and a nearly flat pattern with no obvious changes over the entire time series. The other three indices behaved differently. H$\alpha$ presented a downward trend in both occasions with differing slopes in T1 and T2. The Mg and Na indices had very different patterns from T1 to T2. Both Na and Mg showed an increased, correlated absorption during the second half of the planetary transit duration that extended after the transit in T1. This pattern was, however, not present in T2. The origin of these absorption structures is critical for our study. \citet{llama15,llama16}  simulated the transit of a planet in front of the solar disc and computed a handful light curves corresponding to a few wavelengths in the interval 94--4500\AA{} with the main goal of exploring how the light curves of planetary transits depend on the stellar activity and "orography" at different wavelengths. Indeed, if the planet crosses over stellar active regions, this would cause deviations of the transit curve. But all deviations appear within the transit duration for each wavelength, except for cometary-like atmospheres. One might think that the extra absorption at Na and Mg wavelengths in KELT-7 could be planetary in origin, in which case a similar pattern is expected in all transits. This is not observed in T2 (T2 is significantly different from T1), thus leading us to interpret that the structures seen in the spectral activity indices shown in Fig.~\ref{activity} are due to the star's activity rather than the planet's absorption. 

The {\sl XMM-Newton} EPIC light curve showed that KELT-7 is variable in X-rays in time scales of hours, although no flares were detected during the observations. The OM monitor, more sensitive to chromospheric emission, also showed no flares and, interestingly, a low-level variability ($\la 15\%$). KELT-7's X-ray-to-bolometric luminosity ratio is $\log L_{\rm X}/L_{\rm bol}$~$=$~$-$5.70~$\pm$~0.05. This value is indicative of a level of activity similar to the Sun during the solar maximum \citep[e.g.][]{orl01}. Typically, F2V-type stars have marginal or no detections in X-rays, which contrasts with KELT-7. Its unusually large stellar radius may make it more susceptible to generate convection that may drive some (although still low) activity in the stellar atmosphere.

\subsection{Stellar rotation period}
\begin{figure*}                                      
\centering                       
    \includegraphics[width=0.98\textwidth]{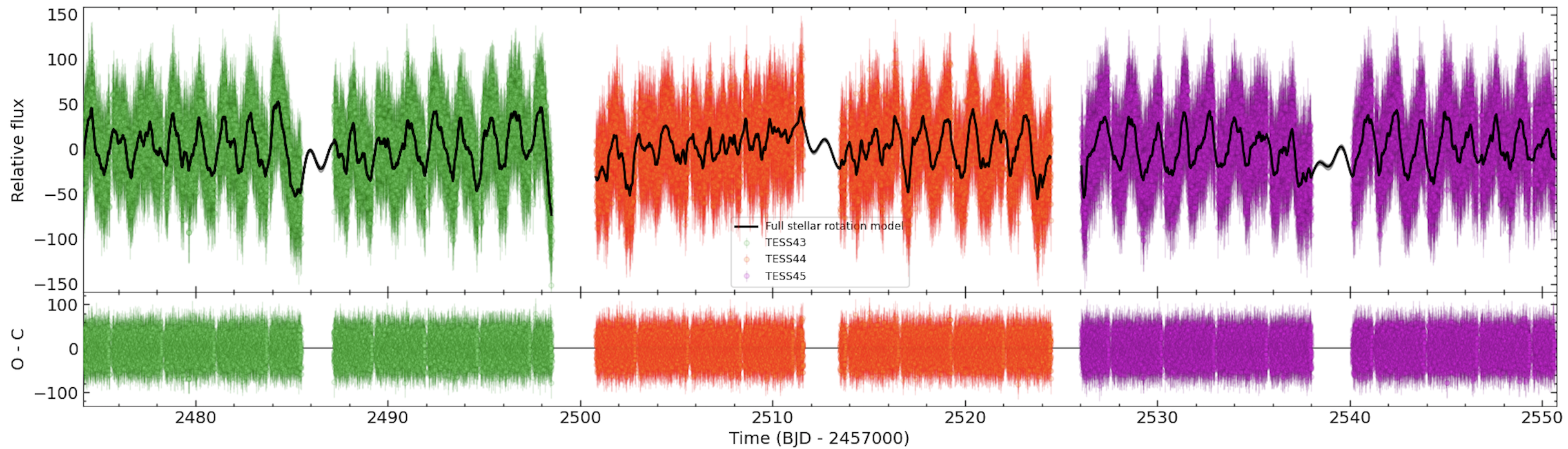}
    \caption{ 
    The PDCSAP fluxes of {\sl TESS} sectors 43 (green), 44 (red), and 45 (magenta) are shown together with the stellar activity model (black) in the top panel. The model yields a stellar rotation period of 1.38~$\pm$~0.05 d and a {\sl TESS} photometric variability amplitude of about 40 e$^-$\,s$^{-1}$ or 450 ppm. The planet transits are removed. The bottom panel illustrates the photometric residuals.
    }
    
    \label{star_rotation}
\end{figure*}

 We used {\sl TESS} Sectors 43, 44, and 45 PDCSAP fluxes observed between 2021 September 16 and 2021 December 2 to determine the star's rotation period and typical photometric amplitude of the variations at optical wavelengths. Sector 19 (from 2019 November 28 through 2019 December 23) was not used because no obvious periodic pattern was seen in the light curve. However, the data of the most recent Sectors show a recurrent structure that is useful for our purposes. We modelled {\sl TESS} Sectors 43--45 light curve by means of Gaussian processes with the {\sl celerite} quasi-periodic kernel or covariance function defined in Eq. 56 of \citet{celerite}, which is recommended to gain in computational efficiency. The stellar rotation period, $P_{\rm {rot}}$, is included in the cyclic component of this kernel; other parameters are the time decay, $L$, of the variations and an "amplitude", $B$, related to the exponential part of the covariance function (there is another parameter, $C$, which we fixed to a null value). We set uninformative, wide uniform priors on $B$ and $L$, while the priors on $P_{\rm {rot}}$ were normal centred at 1.5 d with a width of 0.5 d, which is based on our expectations from the large spectroscopic $v$\,sin\,$i$ stellar value (Section~\ref{sec:stepar}). The transits of KELT-7 b were masked out. We used the {\tt juliet} code \citep{2019MNRAS.490.2262E} for performing the fit of the {\sl celerite} model. In our analysis, the mean flux and the photometric jitter were computed for each Sector independently while the parameters of the covariance function were shared among all three Sectors. The resulting rotation period of KELT-7 is $P_{\rm {rot}}$ = 1.38 $\pm$ 0.05 d, where the error bar corresponds to the 1-$\sigma$ width of the posteriors distribution shown in Fig.~\ref{prot_corner}. We found negligible jitter values for all three Sectors; consequently, they are not shown in Fig.~\ref{prot_corner}. We conducted the same exercise on each Sector separately obtaining stellar rotational periods (1.40 $\pm$ 0.06, 1.38 $\pm$ 0.11, and 1.38 $\pm$ 0.07 d for Sectors 43, 44, and 45, respectively) compatible at the 1-$\sigma$ level, thus supporting the small error bar associated to our $P_{\rm {rot}}$ determination and the lack of significant differential rotation over the 77 d of continuous {\rm TESS} observations. KELT-7's rotation period is listed in Table~\ref{par_planet}. Figure~\ref{star_rotation} displays {\sl TESS} Sectors 43, 44, and 45 light curve together with the best model. The mean flux of each Sector and the planetary transits were conveniently removed for the clarity of the Figure. The periodic stellar variations show a modest amplitude of $\approx$\,40 e$^-$\,s$^{-1}$ or $\approx$\,450 ppm, that is, the peak-to-peak stellar variability, when present in the data, is 9 times less intense than the depth of the planetary transit at optical wavelengths.

The derived fast rotation period of KELT-7 is larger than the expected periodicity assuming a rigid body rotation, a stellar spin-axis inclination angle of 90 deg, and the star's radius and projected rotational velocity determined in Section~\ref{sec:stepar}. The difference is 0.3 d (or $\approx$\,22\,\%). Only by increasing $R_*$ (in line with the scenario of an evolved star) and decreasing $v$\,sin\,$i$ by 4--6\,$\sigma$ do the two periods coincide. This may hint at underestimated error bars of the stellar parameters or the presence of measurable differential rotation. Strong differential rotation appears more frequently among slow rotators while the strength of stellar differential rotation diminishes in stars rotating as rapidly as $v$\,sin\,$i > 50$ km\,s$^{-1}$ \citep{reiners03a,reiners03b}. \citet{2016MNRAS.461..497B} showed that the normalized differential rotation shear, $\Delta \Omega$ (difference in angular velocity between pole and equator), increases to a maximum in F-type stars using {\sl Kepler} data, which agrees with theoretical predictions \citep[e.g.,][]{2012MNRAS.423.3344K}. For stars with the same temperature as KELT-7 and rotation periods in the interval 0.3--2.0 d, \citet{2016MNRAS.461..497B} measured $\Delta \Omega$ = 0.25 rad\,d$^{-1}$ and a relative shear $\alpha = 0.22$. For KELT-7, we determined a differential angular velocity of 0.20 $\pm$ 0.05 rad\,d$^{-1}$, thus suggesting that we are likely viewing this star near equator on, that is, with a stellar spin-axis inclination angle of $\approx$\,90 deg. We will use this result in Section~\ref{sec:RM} for determining the 3D geometry of the KELT-7 planetary system.

\subsection{Planetary parameters}
\label{sec:plapar}

\begin{figure}
\centering
\includegraphics[width=0.5\textwidth]{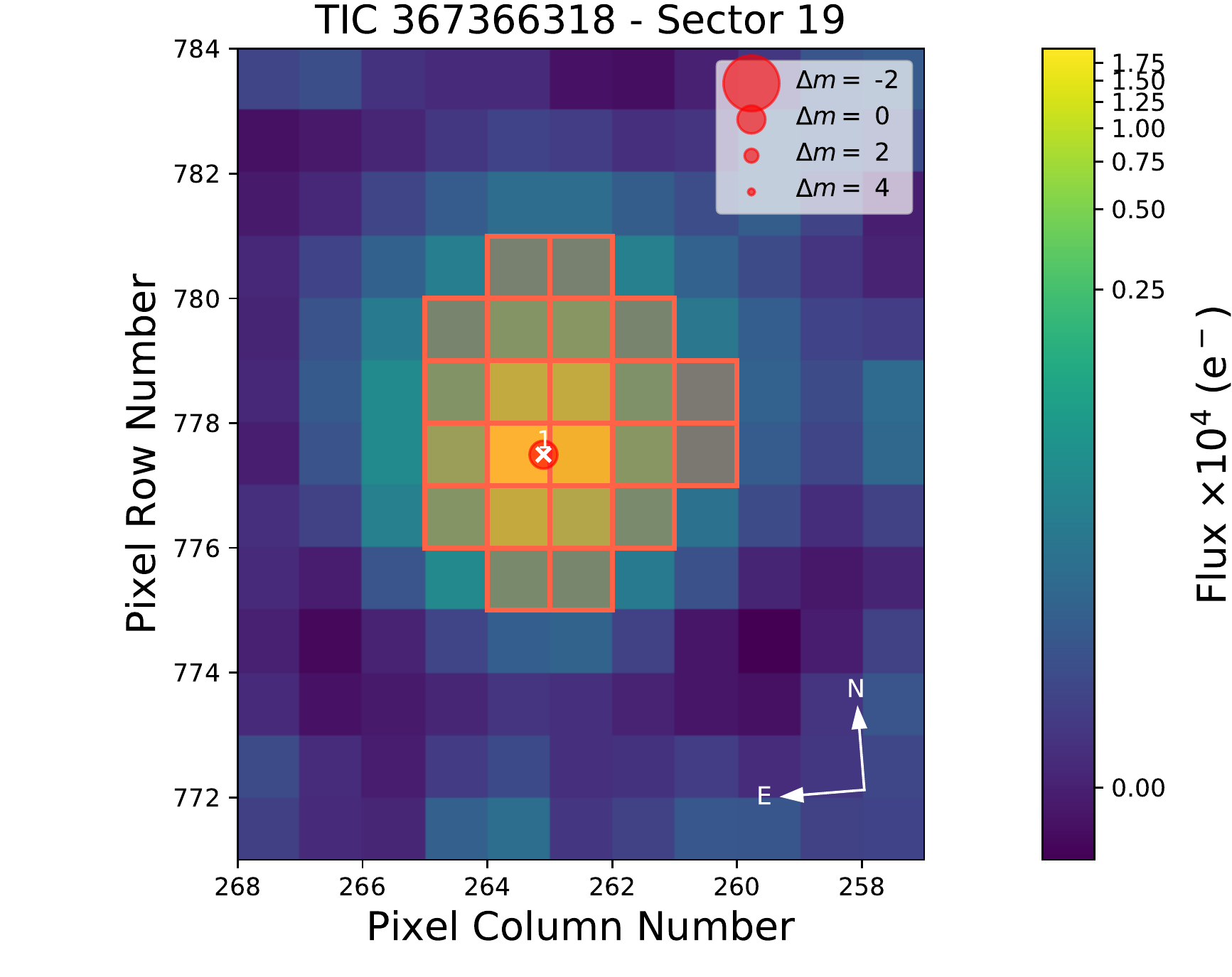}
\caption{Target pixel files (TPF) of KELT-7 (cross symbol) in {\sl TESS} Sector 19. The electron counts are colour-coded. The shadowed pixels correspond to the {\sl TESS} optimal photometric aperture used to obtain the simple aperture photometry (SAP) fluxes. There are not nearby sources that might contaminate the {\sl TESS} light curve with magnitude contrast up to $\Delta m=4$.}
\label{fig:apertures}
\end{figure}

We performed a combined photometric and spectroscopic analysis of the KELT-7 system aimed at improving the available planetary parameters and ephemeris; the latter is necessary for an accurate analysis of the R-M effect and planetary atmosphere. We used {\sl TESS} and KELT photometry and the TRES RVs. Sector 19 of {\sl TESS}, KELT photometry and the TRES RVs were all published in \citet{bier15} and \citet{plu20}.  Here, we also included three new {\sl TESS} Sectors (43--45) containing 24 additional planetary transits that served to improve the accuracy of all light-curve-related planetary parameters. We note that the ephemeris of KELT-7b were also recently refined by \citet{2020AJ....159..137G} using {\sl Spitzer} secondary transits. Our first step was to check that there are no other nearby sources contaminating the {\sl TESS} light curve that could affect the transit analysis. The TPF file of Sector 19 with the standard pipeline apertures is shown in Fig.~\ref{fig:apertures} and it was created with {\tt tpfplotter}\footnote{ \url{https://github.com/jlillo/tpfplotter}} \citep{2020A&A...635A.128A}. The code overplots all sources from the {\sl Gaia} Data Release 2 (DR2) catalogue \citep{GDR2} with magnitude contrast up to $\Delta m=4$ on top of the {\sl TESS} TPFs. No identified sources are within the automatically selected pipeline aperture of KELT-7 where the {\sl TESS} pixel scale is 21\,arcsec\,pixel$^{-1}$. Therefore the extracted light curve is free of any significant contamination.

\begin{figure*}
\centering
\includegraphics[width=\textwidth]{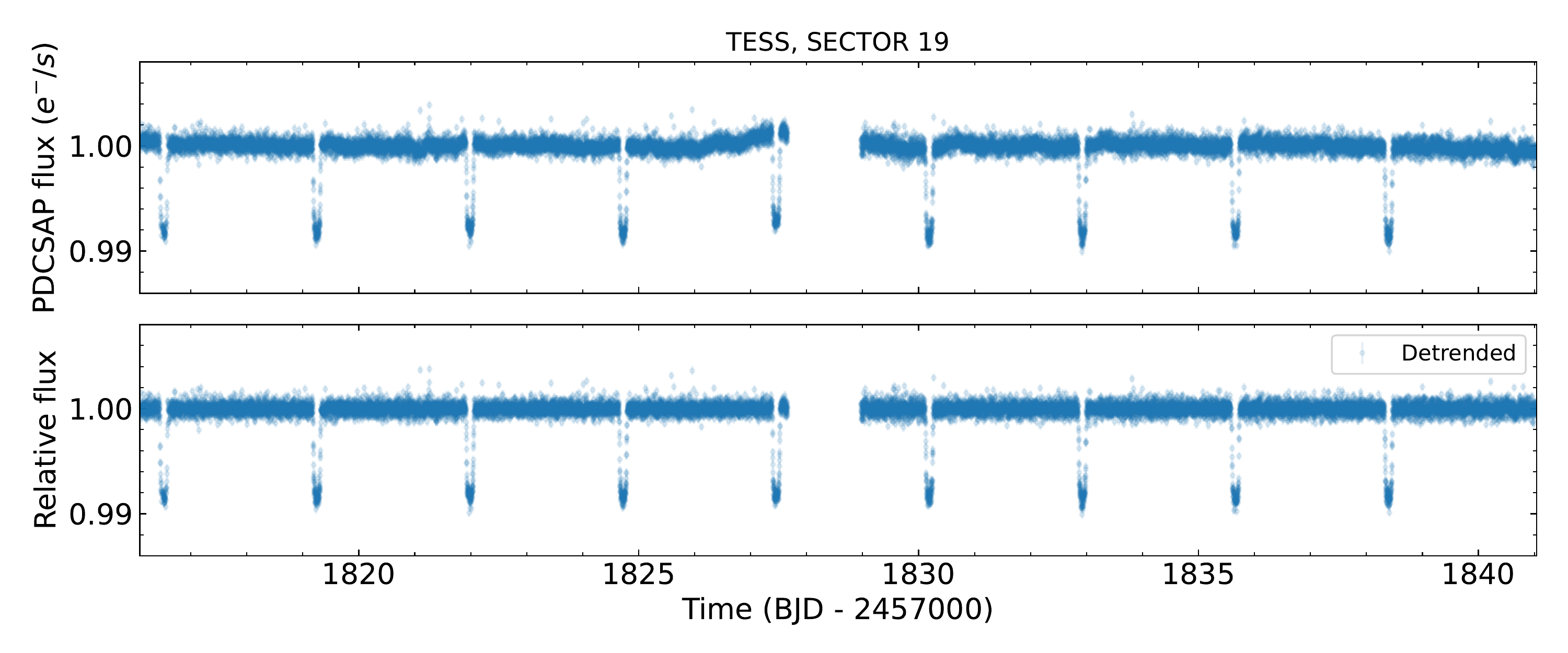}
\caption{KELT-7 \textit{TESS} Sector 19 light curve. {\sl Upper panel:} PDCSAP fluxes. {\sl Lower panel:} PDCSAP fluxes after detrending.}
\label{fig:SAP_and_PDCSAP}
\end{figure*}

In principle, the \textit{TESS} PDCSAP fluxes are already corrected for instrumental variations, but some instrumental residuals and signatures of stellar variability (as shown in the previous Section) may remain. To remove them, we fit the PDCSAP fluxes by using a Gaussian Process (GP) with the {\tt celerite} approximated Matern-3/2 kernel that takes the form:
\begin{equation}
 k_{i,j}(\tau)=\sigma^2\left(1 + \frac{\sqrt{3}\tau}{\rho} \right)\exp{\left( \frac{-\sqrt{3}\tau}{\rho} \right)}
\end{equation}
where $\tau= |t_{i}-t_{j}|$ is the time-lag, $\sigma$ is the amplitude of the GP, and $\rho$ is the time length-scale of the GP. All planetary transits were first masked before the fit so that they do not contribute to the GP model.  We remark that the choice of the covariance function for flattening the light curves has a negligible impact on the results of our paper. We used the {\tt juliet} code \citep{2019MNRAS.490.2262E}, which is based on the {\tt celerite} \citep{celerite} Python package for the GP models, the {\tt batman} package \citep{2015PASP..127.1161K} for the planetary transit light curves, and the {\tt radvel} \citep{2018PASP..130d4504F} package to model Keplerian RV signals. For the flattening of the {\sl TESS} data, we set wide log-uniform and uniform priors on the GP parameters. Both the original PDCSAP fluxes and the resulting detrended light curve {\sl TESS} Sector 19 are shown in Fig.~\ref{fig:SAP_and_PDCSAP}. 

After flattening the {\sl TESS} data, we proceeded to the combined analysis of the planetary system using {\tt juliet}. The limb-darkening effect was taken into account with the parametrisation coefficients $q_1$ and $q_2$ defined by \cite{2013MNRAS.435.2152K} and a quadratic law. We also used the $r_1$ and $r_2$ parametrisation \citep{2019MNRAS.490.2262E} instead of directly determining the relative radii ($p=R_p/R_*$) and the impact parameters ($b$) of the planet: $r_1$ and $r_2$ can vary between 0 and 1 and are defined to explore all the physically meaningful ranges for $p$ and $b$. We set a prior on the stellar density, $\rho$, instead of the scaled semi-major axis of the planet, $a$. As in the literature, we adopted a circular orbit for KELT-7b and defined a normal prior on the orbital period centred at the value from \citet{bier15}. The definitions employed for each individual prior are given in Table~\ref{tab:kelt7_priors_details}. {\tt juliet} was run with a total of 1\,000 live points.

\begin{figure*}
\centering
\includegraphics[width=0.48\textwidth]{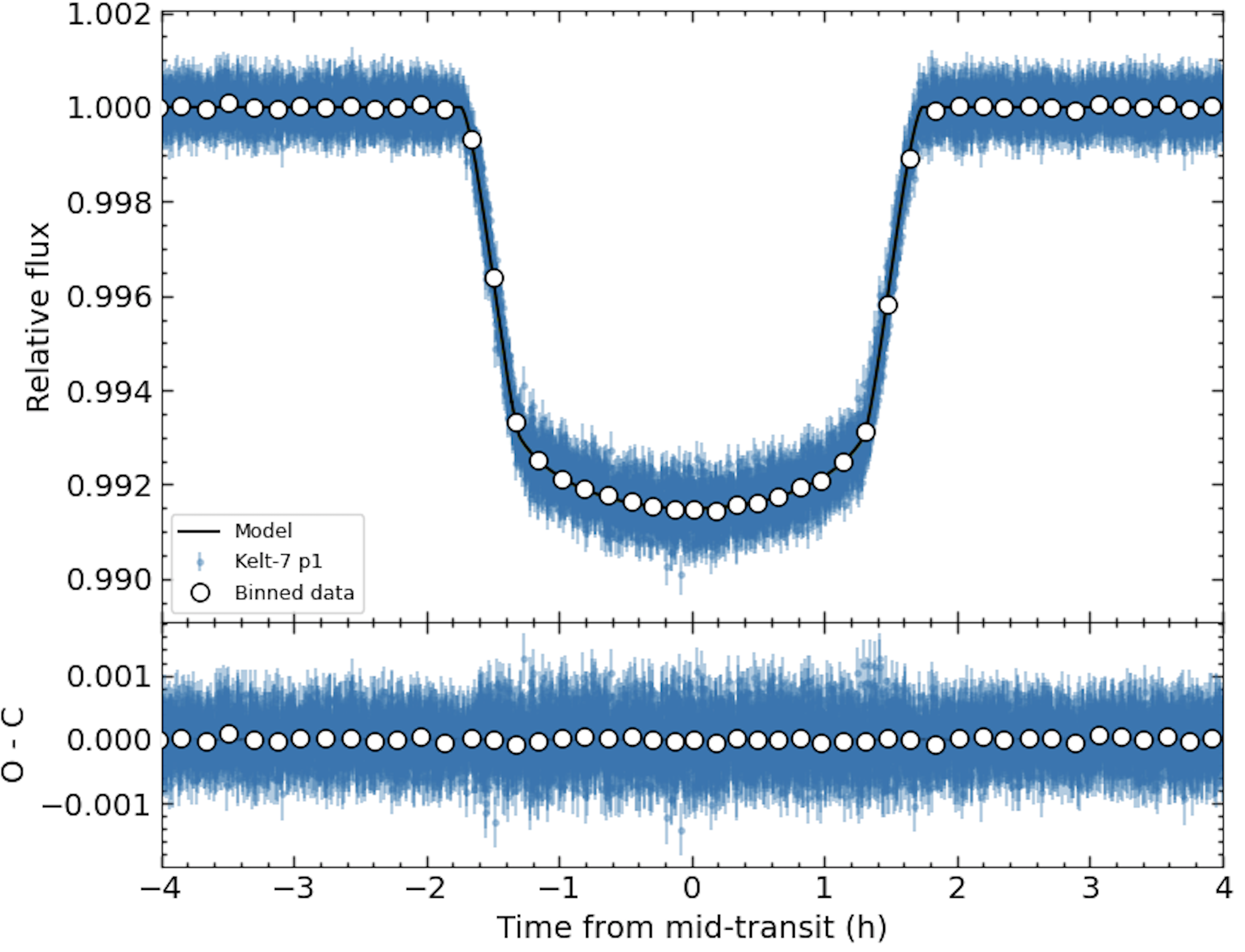}
\includegraphics[width=0.48\textwidth]{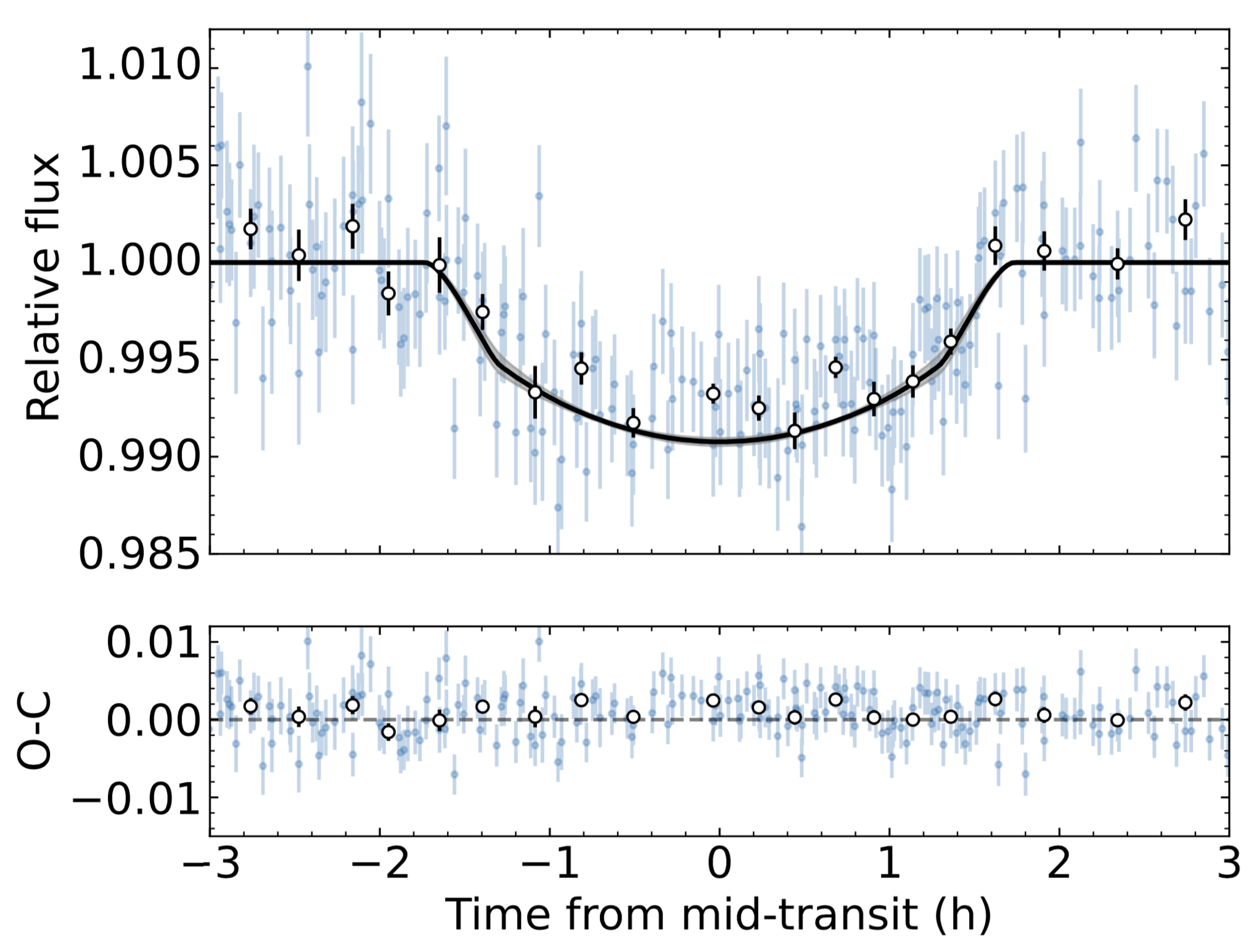}
\caption{KELT-7 b lights curves  ({\sl TESS} on the left and KELT on the right) folded in phase with the planetary period (top panels). The time axis is set to a null value at the mid-transit time. Original measurements are plotted as blue dots, binned data are shown with white dots. The best fit is depicted with a black line. The bottom panels show the residuals of the fit.
}
\label{fig:lc_vs_phase}
\end{figure*}

\begin{figure}
\centering
\includegraphics[width=0.45\textwidth]{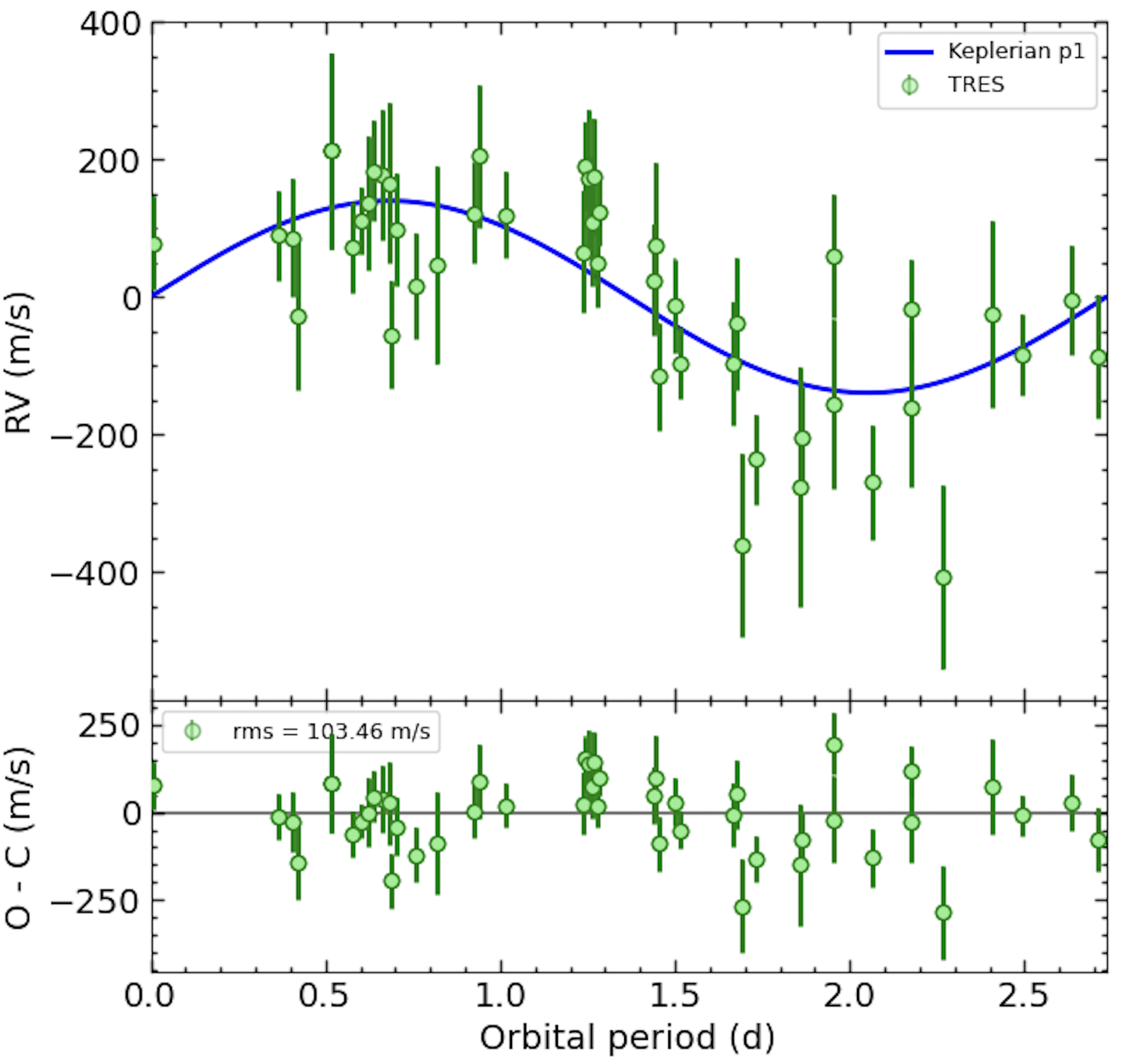}
\caption{ TRES RVs of KELT-7 (green dots) folded in phase with the orbital period of the planet (top panel) together with the best model (blue line). The error bars of the individual RV measurements include the reported uncertainties and the RV jitter resulting from our analysis. The best Keplerian solution has an RV amplitude of 140~$\pm$~17~m~s$^{-1}$. The RVs affected by the R-M effect are not shown. The bottom panels shows the RV residuals.}
\label{fig:rv_vs_phase}
\end{figure}

The results of our analysis are listed in Table~\ref{par_planet}; Figure~\ref{fig:kelt7_cornerplot} displays the distributions of the posteriors in a corner plot. The improvement on the determination of the periastron passage ($T_0$) and orbital period ($P$) is about a factor of 10 with respect to the values reported in the discovery paper and is similar to those reported by \citet{2020AJ....159..137G} and \citet{plu20}. The {\sl TESS} and KELT light curves folded in phase with the planetary period and the best fit are illustrated in Figure \ref{fig:lc_vs_phase} while the RVs are shown in Figure~\ref{fig:rv_vs_phase}. The $rms$ of the RV residuals is 103.5~m\,s$^{-1}$, which is about 1.5 times smaller than the $rms$ of the \citet{bier15} solution and closer to the quoted TRES RV error bars

The planet KELT-7\,b has a mass of $\rm 1.28 \pm 0.17\,M_{\rm Jup}$, an inflated radius of $\rm 1.496 \pm 0.035\,R_{\rm Jup}$ and a mean density of $\rm 0.48 \pm 0.10\,g~cm ^{-3}$; it is located at a separation of $0.0434 \pm 0.0012$\, au from its host star and orbits around the centre of mass of the system with a period of $2.73476550 \pm 0.00000033$\,d. Finally, we derived the equilibrium temperature ($T_{\rm eq}$) of 2028~$\pm$~17~K thanks to the Stefan–Boltzmann equation, the stellar and planetary parameters given in Table~\ref{par_planet}, and assuming a  null planetary albedo (A = 0.0).

\section{Rossiter-McLaughlin effect}
\label{sec:RM}
\subsection{Classical Rossiter-McLaughlin effect analysis}
\label{sec:classrm}

\begin{figure}                                      
\centering                       
    \includegraphics[width=0.5\textwidth]{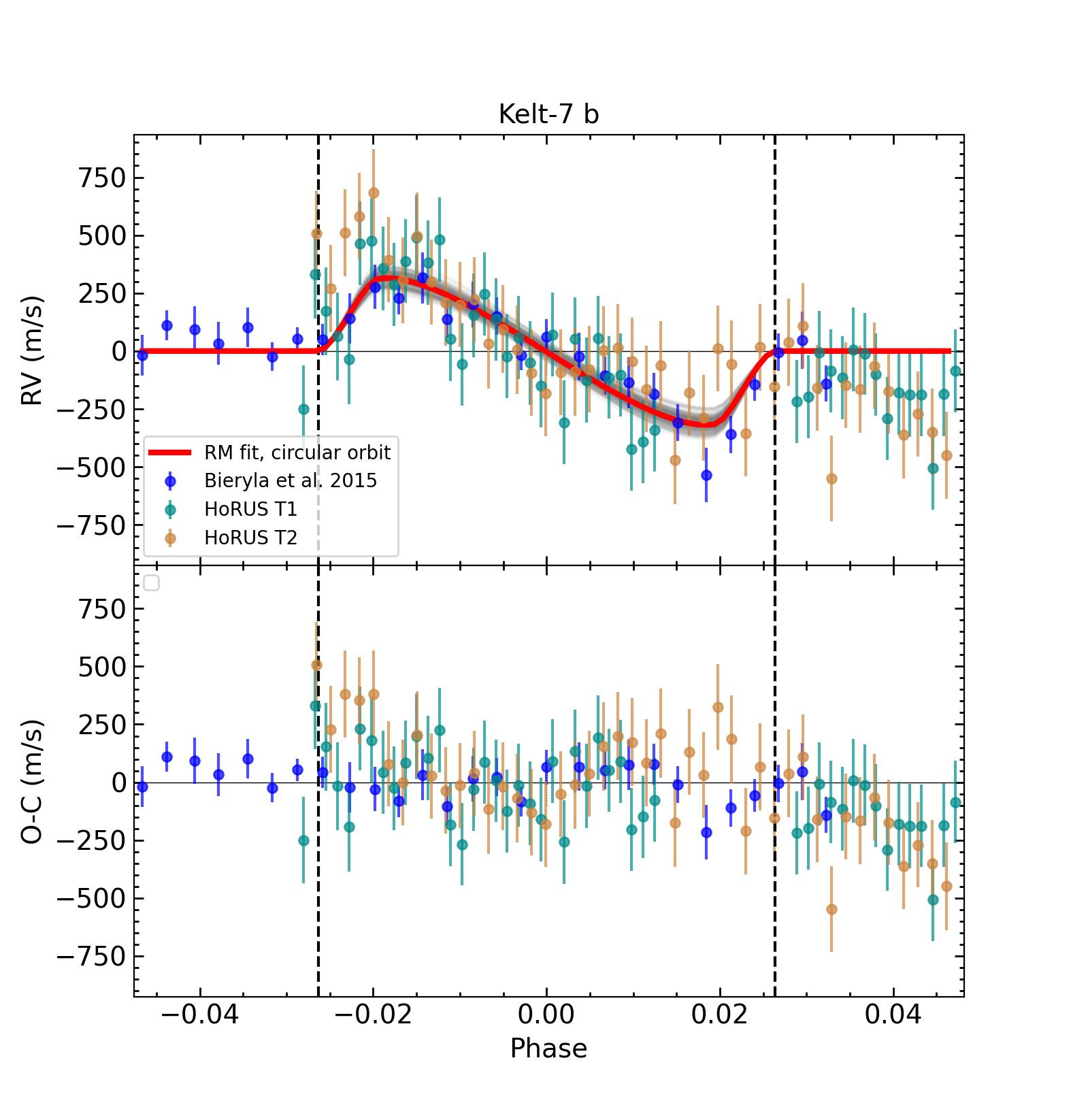}
    \caption{Rossiter-McLaughlin effect of KELT-7b as a function of the planetary orbital phase (top panel). The best fit and its associated uncertainty are shown with a red line and a greyish area, respectively. The bottom panel shows the RV residuals with an $rms$ of the same size as the individual error bars.}
    \label{rossiter_fit}
\end{figure}

We used the refined ephemeris of KELT-7b derived in Sect.~\ref{sec:plapar} and the prescription of \citet{2013A&A...550A..53B} to analyse the R-M curve and obtain the projected spin-orbit angle between the stellar spin axis and the transiting planet's orbital axis. The formulas in \citet{2013A&A...550A..53B}, implemented in the ARoME code\footnote{\url{http://www.astro.up.pt/resources/arome}}, are adequate for RVs obtained using the CCF method. We combined a total of three transits: that of \citet{bier15} and the two HORuS ones. The velocity amplitude of the R-M effect appeared to be constant in all three given the relatively large error bars of the measurements as it is illustrated in Fig.~\ref{rossiter_fit}. We first removed the Keplerian signal from the RVs and phase folded all data using the revised planet orbital period and $T_0$ parameters. To run the ARoME code, we obtained the limb-darkening coefficients for KELT-7 and the wavelength coverage of HORuS using  the Exoplanet Characterization toolkit\footnote{\url{https://exoctk.readthedocs.io/en/latest/}}. This toolkit contains a limb darkening calculator\footnote{\url{https://exoctk.stsci.edu/limb\_darkening}}, which delivers $c_1 = 0.393 \pm 0.014$ and $c_2 = 0.270 \pm 0.018$ for the quadratic limb-darkening profile. We also adopted the spectral resolution of HORuS (12~km\,s$^{-1}$) as the width of the non-rotating star, the quantity of 71.4~km\,s$^{-1}$  as the width of a rotating star measured from the HORuS data, a macro turbulent velocity of 14~km\,s$^{-1}$ determined for KELT-7's temperature and high surface gravity from the study of \citet{brewer16}, and the planetary and stellar parameters of Table~\ref{par_planet}. Therefore, two free parameters remained: the projected obliquity angle, $\lambda$, and the stellar projected rotational velocity.

We ran 4 parallel chains, each consisting of 2\,000 iterations. The first 1\,000 iterations were discarded as burn-ins to allow for reasonable mixing and  infer $\lambda$ and its corresponding uncertainty. All priors are wide and uniform ($\mathcal{U}$($-90, 90$) deg and $\mathcal{U}$(0.1, 120) km\,s$^{-1}$ for $\lambda$ and $v$\,sin\,$i$, respectively). The result indicates that the sky-projected spin-orbit alignment is $\lambda$~$=$~0.5\,$^{+2.4}_{-2.9}$~deg. Our obliquity angle agrees within 2 $\sigma$ with the value reported in \cite{bier15}, although our measurement is improved by a factor of two given the larger number of RVs (transits) in our analysis. The best fit model and its associated uncertainties are depicted in Figure~\ref{rossiter_fit}. The distribution of the posteriors, which are quite Gaussian-like, is given in Fig.~\ref{rm_corner}. Our derivation of the stellar rotation velocity using the ARoME code is 38~$\pm$~3~km\,s$^{-1}$, which is significantly smaller than the spectroscopic measurement obtained in Section~\ref{sec:stepar}. This anomaly was already reported by \citet{2017MNRAS.464..810B}, who found that the \citet{2013A&A...550A..53B} model clearly tends to underestimate the value of $\varv \sin{i}$ compared to the spectral analysis by a fractional difference of about 10--80\,\% (see Fig.~31 of \citealt{2017MNRAS.464..810B}). In fact, the work of \citet{2017MNRAS.464..810B} explored the underlying cause of this anomaly (i.e. differential rotation, convective blueshifts). Their findings suggest that differential rotation is unlikely to explain the underestimated $\varv \sin{i}$ and that convective blueshifts are not important for fast-rotators. For fast-rotating stars, the relation between measured RV and the occultation of the local stellar surface by the planet is not straightforward. Also, the approximations made in \citet{2013A&A...550A..53B} might not be fully valid for high rotation velocities. We provide a more reliable derivation of the projected obliquity in the next Section by studying the Doppler Shadow. 

\subsection{Doppler Tomography}
\label{doptom}

The amplitude of the R-M effect during the transit of KELT-7b is $\approx$300~m~s$^{-1}$ due to the high $\varv \sin{i}$ of its parent star. During the transit, the planet covers different regions of the stellar disc producing a Gaussian bump on the stellar CCFs \citep[see, e.g.][]{col10,ceg16}. This Gaussian bump is also known as the Doppler shadow and corresponds to those regions of the stellar disc covered by the transiting planet. Consequently, we decided to explore the presence of the Doppler shadow corresponding to KELT-7b in our data. 

\begin{figure*}                                      
\centering                       
   \includegraphics[width=\textwidth]{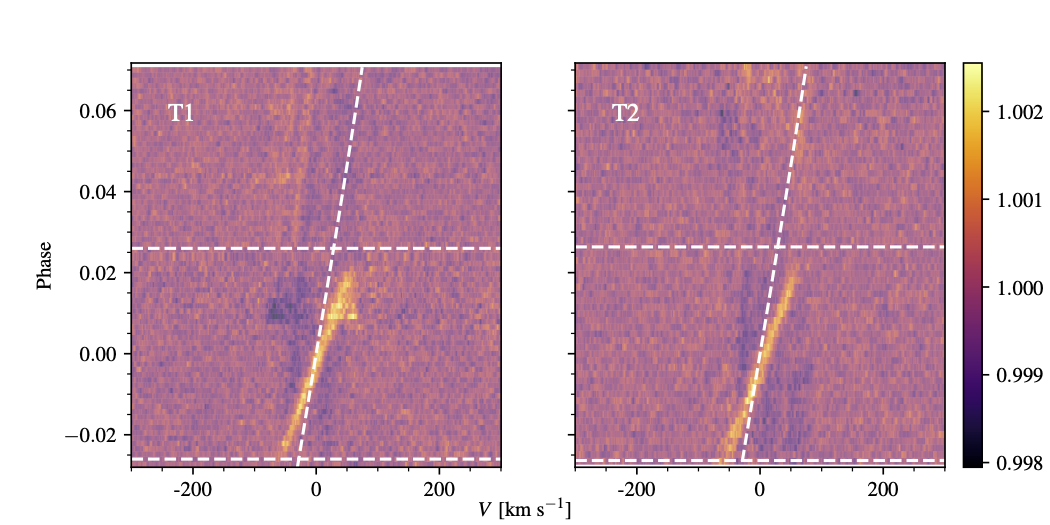}
    \caption{Doppler tomography map for the two HORuS transits of KELT-7b. The transit window is defined by the two horizontal dashed lines whereas the planetary Keplerian path is represented by a dashed (inclined) curve.}
    \label{tomo}
\end{figure*}

In order to recover the Doppler shadow from the data, we need to remove the stellar contribution from the CCFs. To that aim, we gathered the CCFs computed in Sect~\ref{radvel} and scaled them to the same flux level dividing by a first-order polynomial. Then, we classified the CCFs into two different categories: in-transit and out-of-transit. We created a master CCF combining only the out-of-transit CCFs.  We divided the data by this master CCF in order to reveal the signal imprinted by the planet in the CCFs. We used the resulting residuals to produce the Doppler line tomography map depicted in Fig.~\ref{tomo}. The resulting map should in theory contain three different strong signals, the R-M effect, the stellar centre-to-limb variation, and the planet atmosphere, although typically at different velocities. In fact, most of the metallic lines contained in the stellar mask used by us are due to iron, cobalt, and nickel \citep [see, e.g.][]{ehr20}. If absorption by these species is present in the atmosphere of KELT-7b, we would expect to see a signal in the 2D map of Fig.~\ref{tomo} following the path of the planetary Keplerian velocity. As illustrated in Fig.~\ref{tomo}, only the R-M signal is clear, which implies that KELT-7b has no neutral iron/cobalt/nickel in gaseous form or the planetary signature lies below our detectability limit. 

Many UHJ planets have neutral and ionised iron in their atmospheres (see Section~\ref{sec:intro}); KELT-7b has a cooler equilibrium temperature than the so-called UHJs and lies near the temperature borderline that separates the detection and non-detection of neutral iron in the planetary atmospheres. WASP-19b, a hot Jupiter of similar mass and equilibrium temperature as KELT-7b, has titanium oxide and water vapour in its atmosphere \citep{2017Natur.549..238S,sed21}, which are clearly indicating cool atmospheric layers where a great abundance of neutral atomic iron is not expected.

Finally, given the strength of the Doppler Shadow in the tomography plots, we employed the model described by \citet{col10} to obtain an alternative estimation of the $\varv \sin{i}$ and $\lambda$ independently for T1 and T2. We employed the ephemeris calculated in Sect.~\ref{sec:plapar} and obtained the limb-darkening coefficient for KELT-7 and the wavelength coverage of HORuS using the limb darkening calculator tool\footnote{\url{https://exoctk.stsci.edu/limb\_darkening}}, which delivers $c = 0.601 \pm 0.007$ for the linear limb-darkening profile. Then, we used the {\tt emcee} Python package \citep{emcee}. We ran 20 parallel chains, each consisting of 20\,000 iterations. The first 2\,000 iterations were discarded as burn-ins to allow for reasonable mixing and retrieve $\lambda$, and $\varv \sin{i}$ and their corresponding uncertainties. The result indicates that the sky-projected spin-orbit alignment is for T1 is $\lambda$~$=$~$-$10.13~$\pm$~0.33~deg, whereas T2 yields a value of $\lambda$~$=$~$-$11.38$^{+0.46}_{-0.45}$~deg. The distribution of the posteriors corresponding to both T1 and T2, all of which are well-defined and  Gaussian-like, are given in Fig.~\ref{dopsha_corner}. Our derivation of the stellar rotation by directly fitting the Doppler shadow is 71.87~$^{+0.69}_{-0.71}$~km\,s$^{-1}$ for T1 and 71.21~$^{+0.84}_{-0.83}$ km\,~s$^{-1}$ for T2. We averaged the results corresponding to both transits and we retrieve the following values: $\varv \sin{i}$~$=$~71.87~$\pm$~0.54~km\,s$^{-1}$ and $\lambda$~$=$~$-$10.55~$\pm$~0.27~deg. The derived $\varv \sin{i}$ value is in agreement with the spectroscopic measurement obtained in Section~\ref{sec:stepar}.

Using Eq$.$ 2 of \citet{albrecht21}, we determined the 3D obliquity of the KELT-7 planetary system to be $\psi$ = 12.4 $\pm$ 11.7 deg (or 167.6 deg), where the error bar accounts for the uncertainties in the sky projected obliquity and the orbital inclination angle. We adopted a generous error bar of $\pm$15 deg for the stellar spin-axis angle. KELT-7 is an aligned system. The planet KELT-7b adds to the increasing population of giant planets with perpendicular orbits. 

 \section{Transmission spectroscopy}
 \label{sec:trans}

\subsection{Extraction of the planetary atmosphere} 
\label{sec:exatm}

Prior to exploring the atmosphere of KELT-7b using the HORuS spectra, we corrected the data for telluric absorption by means of the logarithmic scaling relations of line intensity with air mass \citep{sne08,vid10,ast13}. Using the telluric corrected spectra, we extracted the transmission spectrum of KELT-7b via the technique described in \citet{wyt15}. First, we shifted the spectra to the stellar reference frame by subtracting the stellar reflex motion using the data of Table~\ref{par_planet}. Second, each individual observation was flux-scaled by means of a first order polynomial to the lowest air mass spectrum in its corresponding night. Then, we separated the HORuS observations into two groups: in-transit and out-of-transit spectra. The out-of-transit spectra were combined into a master spectrum that we used to remove the stellar contribution from the data. After removing the stellar contribution we ended up with a set of residual spectra that should contain the planetary information. The residual spectra were then shifted in velocity to the planet rest frame using the Keplerian velocities of KELT-7b for each observing time.

We explored the presence of the following atomic species: \ion{H}{i} (6562.78~\AA), \ion{Li}{i} (6103.65~\AA, the resonance line at 6707.82~\AA{} is not covered by HORuS), \ion{Na}{i} (5889.95 and 5895.92~\AA{}), \ion{Mg}{i} (5167.32, 5172.68~\AA{}), and \ion{Ca}{ii} (3933.66 and 3968.47~\AA{}) independently for T1 and T2. The middle and right panels of Figures~\ref{map_Na}--\ref{map_Ca} display the 2D maps around these wavelengths at the velocity rest frame of the star. The planetary 1D transmission spectra at the planet rest frame are illustrated in the bottom middle and right panels. The extracted planetary spectra have an average $rms$ of 0.150\% for the \ion{Na}{i}~D lines, 1.620\% for H$\alpha$, 0.530\% for \ion{Ca}{ii}~H\&K, 0.163\% for \ion{Mg}{i}~b, and 0.187\% for \ion{Li}{i}. Both observed transits have a similar quality; we thus combined them without applying any specific weight. The 2D maps displaying the \ion{Na}{i} and \ion{Mg}{i} wavelength regions (see Figs.~\ref{map_Na} and \ref{map_Mg}) contain faint structures that do not follow the velocity trace delineated by the giant planet. The 2D maps of H$\alpha$ (Fig.~\ref{map_Ha}) and \ion{Ca}{ii}~H\&K (Fig.~\ref{map_Ca}), on the other hand, have larger noise given the strong stellar lines. The map of the \ion{Li}{i} region (Fig.~\ref{map_Li}) does not show any feature and appears rather "flat".

\begin{table}
\caption{Limits to the presence of the species studied in this work. The values provided here correspond to the average of the two transits.}            
\centering
\label{rms_compare}      
\begin{tabular}{lcccc}   
\hline
Line(s)  &      \multicolumn{3}{c}{$rms$}         \\  
               &   R-M~$+$~CLV  & activity &  Observed  limit & Corrected limit    \\  
               &   [\%]   & [\%]  & [\%] & [\%] \\
\hline      
\ion{Ca}{ii}~H\&K & 0.089  & 0.081 &  0.530 & 0.516\\
\ion{Mg}{i}~b  & 0.073  & 0.054 &  0.150 & 0.119\\
\ion{Na}{i}~D  & 0.080  & 0.120 &  0.163 & 0.076\\
\ion{Li}{i}  & 0.182  & --    &  0.187 & 0.043 \\
H$\alpha$   & 0.228  & 0.782 &  1.620 & 1.400\\
\hline 
\end{tabular}
\end{table}

\subsection{Transit effects (R-M+CLV)}
\label{sec:rmpclv}
The extracted planetary data may be contaminated by the R-M effect, the centre-to-limb variation (CLV), and stellar activity. These effects may hide the planet signal at varying degrees depending on the geometry of the planetary system \citep[see, e.g.][]{cze15,yan18,bor18,cas20} and stellar activity changing patterns within the transit duration. We modelled both the CLV and the R-M effect following the approach of \citet{yan18}. The stellar disc is divided into small square pieces with a size of 0.01~$R_{*}$. We assigned to each piece a $\mu$ value that is defined as $\mu$~$=$~$\cos{\theta}$, where $\theta$ is the angle between the normal to the stellar surface and the line of sight. Then, we computed a grid of synthetic model spectra at different $\mu$ values from 0.0 to 1.0 taking a step of 0.05 using the {\tt Turbospectrum}\footnote{\url{https://github.com/bertrandplez/Turbospectrum2019/}} radiative transfer code \citep{turbospectrum} with an interpolated PHOENIX model corresponding to the stellar atmospheric parameters given in Table~\ref{par_planet}. Then, we simulated the planetary transit by blocking those parts of the stellar disc covered by the planet according to the planet ephemeris. The final step was to integrate over the stellar disc to produce a model of the planet passing in front of its parent star. This process allowed us to quantify the contribution of the R-M and CLV to our extracted planetary spectrum that we show in the leftmost panels in Figs.~\ref{map_Na}--\ref{map_Ca}. These simulations show that the R-M and CLV contributions to the 1D transmission spectrum of KELT-7b have an $rms$ of  0.080\% for the \ion{Na}{i}~D lines, 0.228\% for H$\alpha$, 0.089\% for \ion{Ca}{ii}~H\&K, 0.073\% for \ion{Mg}{i}~b, and 0.182\% for \ion{Li}{i}. These theoretical values are smaller than the $rms$ in the transmission spectra. Therefore, there are other pollution sources that are not taken into account in the R-M+CLV model and are likely due to stellar activity.

\subsection{Impact of stellar variability}
\label{sec:stelact}

We found that the transmission spectrum of KELT-7b around the sodium doublet changes from blue-shifted absorption in T1 to emission in T2 (Fig.~\ref{map_Na}): this is  an entirely unexpected behaviour that led us to split each transit in two halves and investigate the temporal evolution of this spectral feature. The top panels of Figure~\ref{Na_activity} show the collapsed planetary spectra for each half transit per observing epoch. In T1, the first half transit has emission-like features while the second half transit shows absorption-like features.  This pattern correlates with the structure of the stellar Na index, also in T1, displayed in Fig.~\ref{activity}, which  prompted us to investigate further the impact of the stellar variability, although small, in our extracted planetary spectra. To this aim, we modelled \ion{Na}{i} line profiles with the width of our observations and depths calibrated to reproduce the variations of the spectroscopic Na index (Sect.~\ref{sec:activity}) for each observing time. We included some noise in the theoretical spectra to mimic real observations. No planetary signal is added to these models. Then, we applied the technique of \citet{wyt15}, including the wavelength shift to bring all data to the planetary velocity rest frame, to collapse the simulated spectra. What we obtained is illustrated in the bottom panels of Fig.~\ref{Na_activity} for both T1 and T2. We were able to reproduce both the shape and intensity of the observed features in the planetary spectrum. The resemblance between the observations and the computations is remarkable and clearly indicates that the emission- and absorption-like features are due to stellar variability and have no planetary origin. We caution that stellar changes occurring during planetary transit observations (including out-of-transit data)  might produce fake planetary detections.

We repeated the same procedure for H$\alpha$, \ion{Mg}{i} b, and \ion{Ca}{ii} H\&K finding the same result (Figs.~\ref{trans_features_Ha}-- \ref{trans_features_Ca}). The relative variations of the stellar activity at these wavelengths during the HORuS observations is dominating the extracted planetary transmission spectrum. Even small changes of $\approx$1\,\%~in the stellar activity indices produce strong effects in the planetary spectrum. \citet{2020A&A...639A..49G} reported night-to-night variations in the \ion{He}{i} absorption signal of HD\,189733\,b detected with the transmission spectroscopy technique. These authors also monitored the $H\alpha$ and other chromospheric lines of the stellar spectrum and concluded that the variability in the planetary helium absorption signal is likely due to variations in stellar activity.

\subsection{Upper limits}
\label{sec:ulims}

\begin{figure*}                                      
\centering                       
   \includegraphics[width=\textwidth]{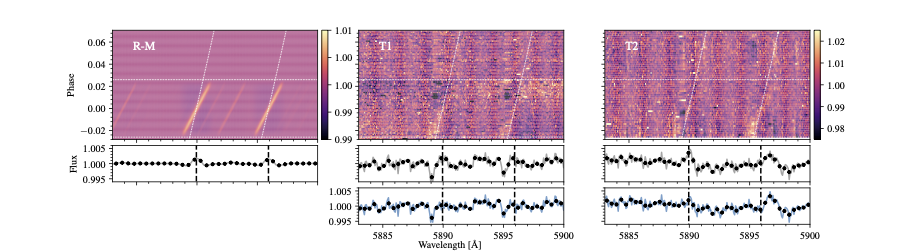}
    \caption{Left panel: modelled R-M and CLV effects for the \ion{Na}{i} doublet. Middle and right panels: KELT-7 \ion{Na}{i} doublet tomography of HORuS T1 and T2 observations. The transit window calculated with the revised ephemeris of Table~\ref{par_planet} is marked by the two horizontal dashed lines, whereas the planet velocity trace is drawn by the two vertical, inclined curves. All tomography 2D maps are in the velocity rest frame of the star. Bottom panels: The grey line represents the stacked planetary transmission spectrum of KELT-7b (planetary rest frame) whereas the black dots stand for a 5-pixel binning of the data. The blue lines represent the stacked planetary transmission spectrum of KELT-7b after the removal of the R-M+CLV model. The position of the \ion{Na}{i} lines are indicated by two vertical dashed lines.}
    \label{map_Na}
\end{figure*}

\begin{figure*}                                      
\centering                       
   \includegraphics[width=\textwidth]{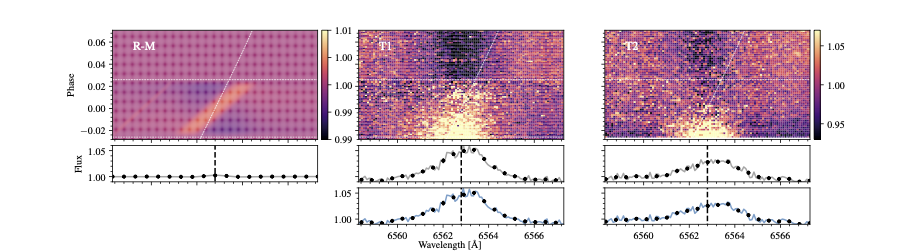}
    \caption{Same as Fig.~\ref{map_Na} but for H$\alpha$.}
    \label{map_Ha}
\end{figure*}

\begin{figure*}                                      
\centering                       
   \includegraphics[width=\textwidth]{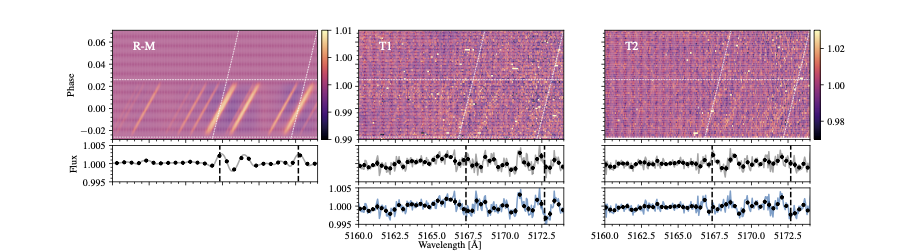}
    \caption{Same as Fig.~\ref{map_Na} but for the \ion{Mg}{i}~b lines.}
    \label{map_Mg}
\end{figure*}

\begin{figure*}                                      
\centering                       
   \includegraphics[width=\textwidth]{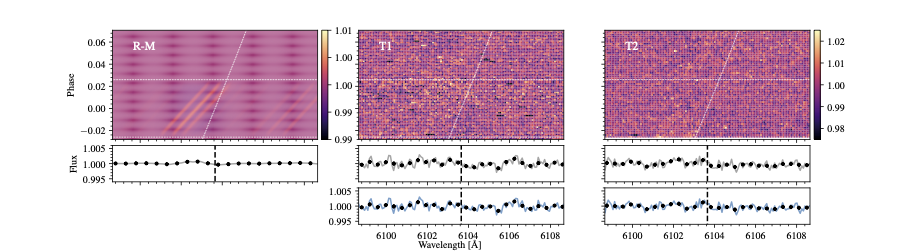}
    \caption{Same as Fig.~\ref{map_Na} but for the \ion{Li}{i} line at 6103.65~\AA{}.}
    \label{map_Li}
\end{figure*}

\begin{figure*}                                      
\centering                       
   \includegraphics[width=\textwidth]{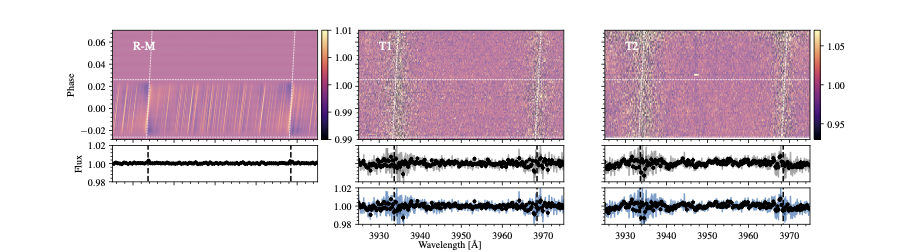}
    \caption{Same as Fig.~\ref{map_Na} but for the \ion{Ca}{ii}~H\&K lines.}
    \label{map_Ca}
\end{figure*}

\begin{figure*}                                      
\centering                       
   \centerline{ \includegraphics[width=0.5\textwidth]{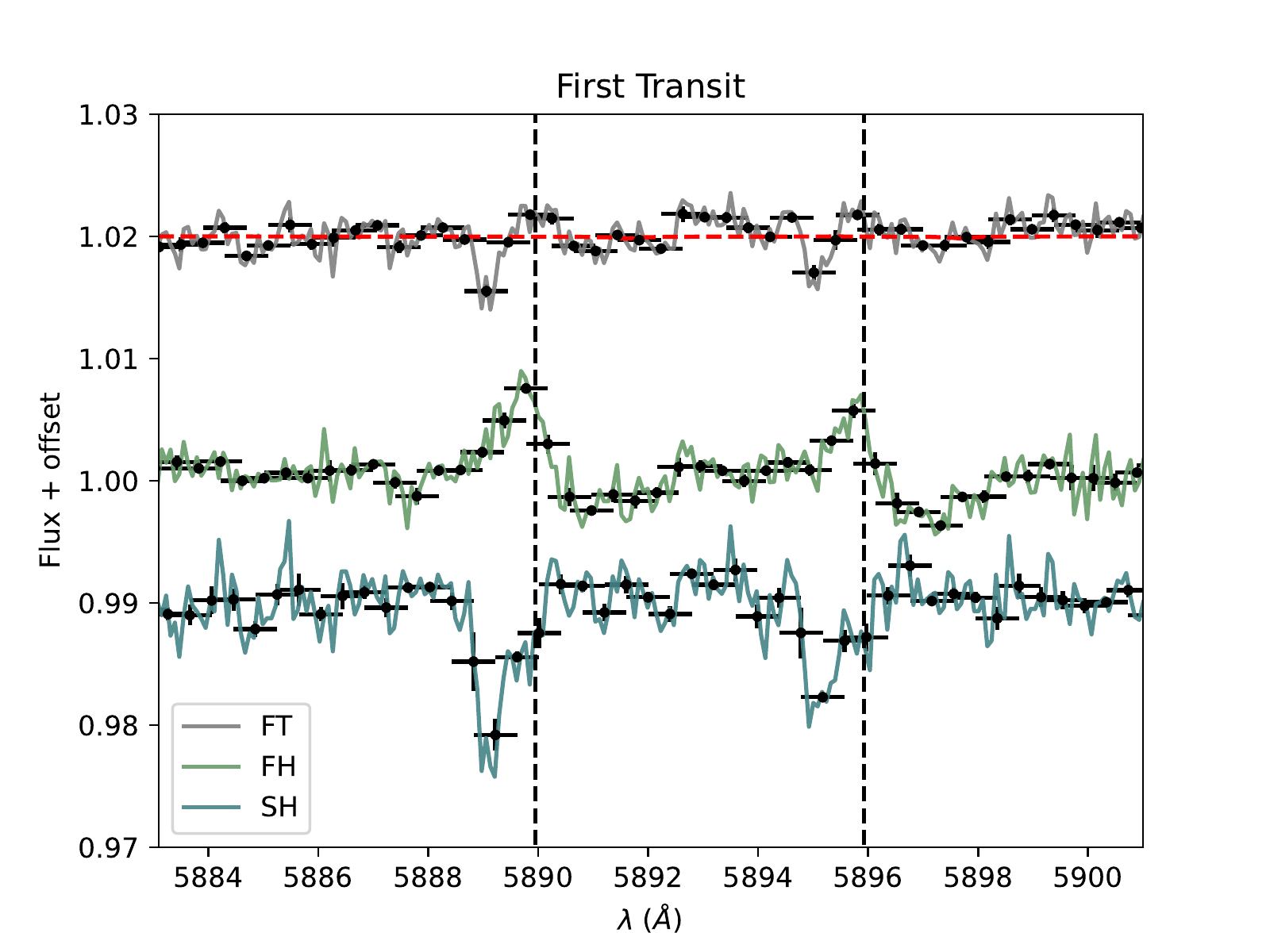}
    \includegraphics[width=0.5\textwidth]{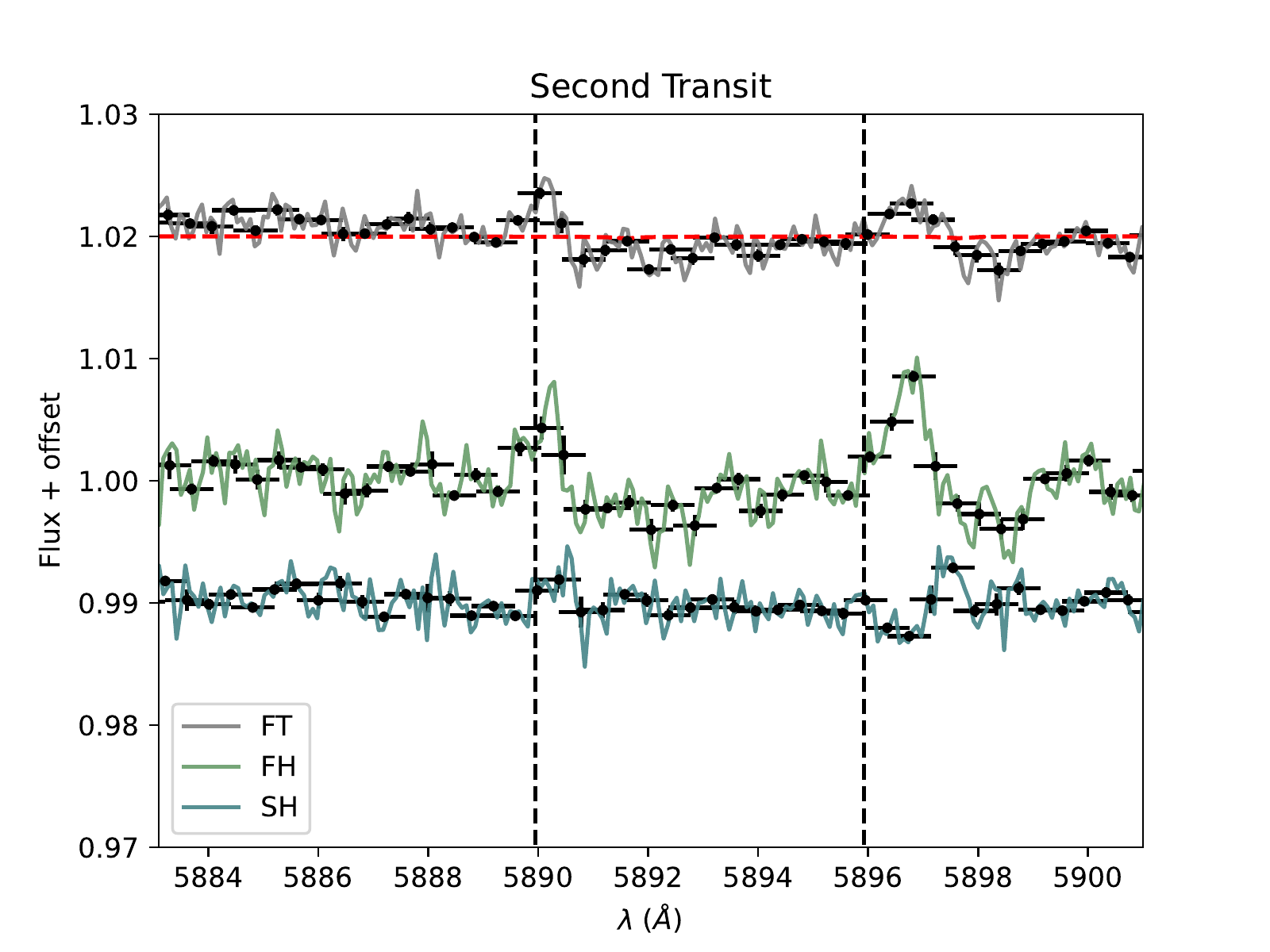}}
       \centerline{ \includegraphics[width=0.5\textwidth]{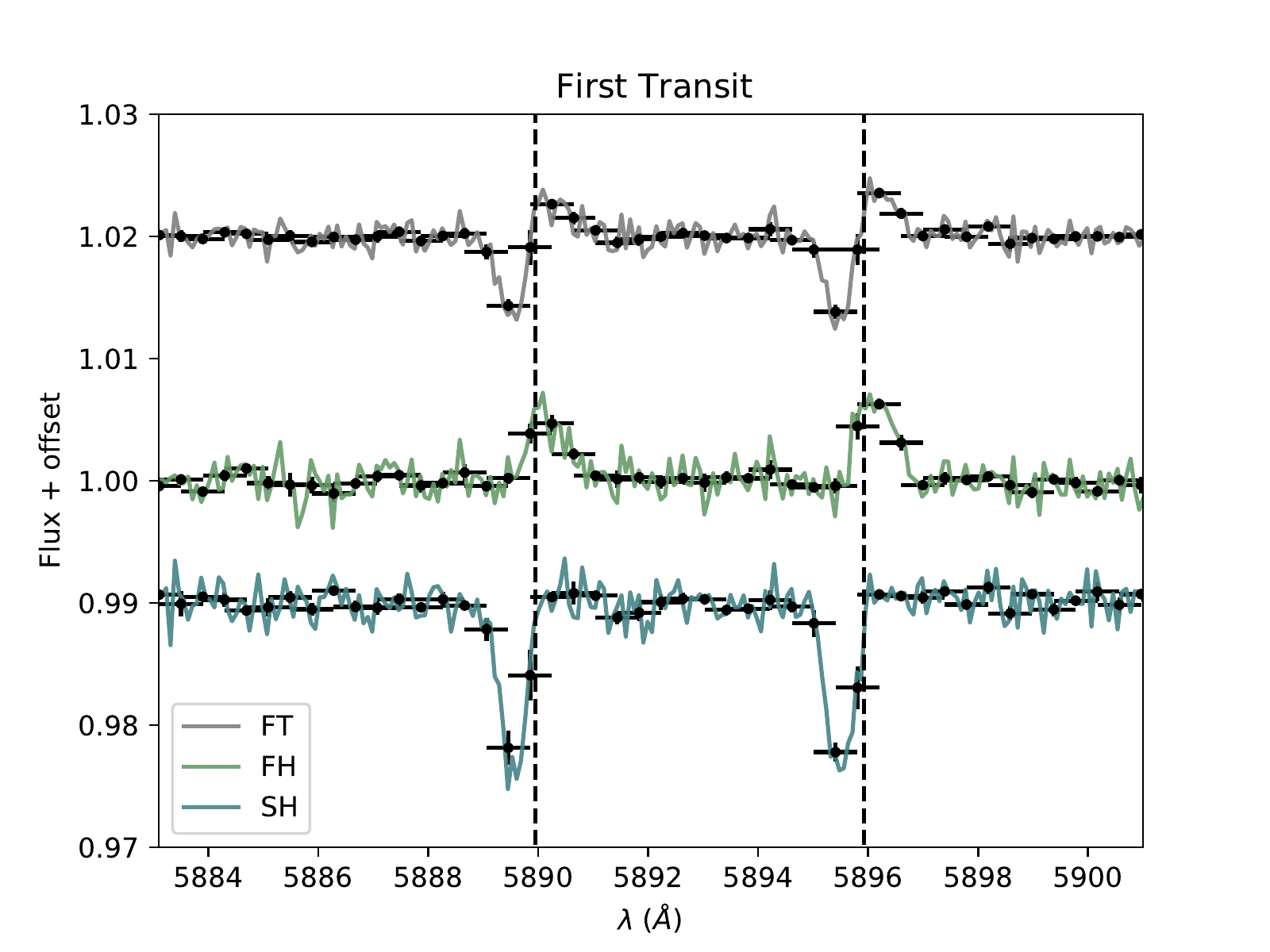}
    \includegraphics[width=0.5\textwidth]{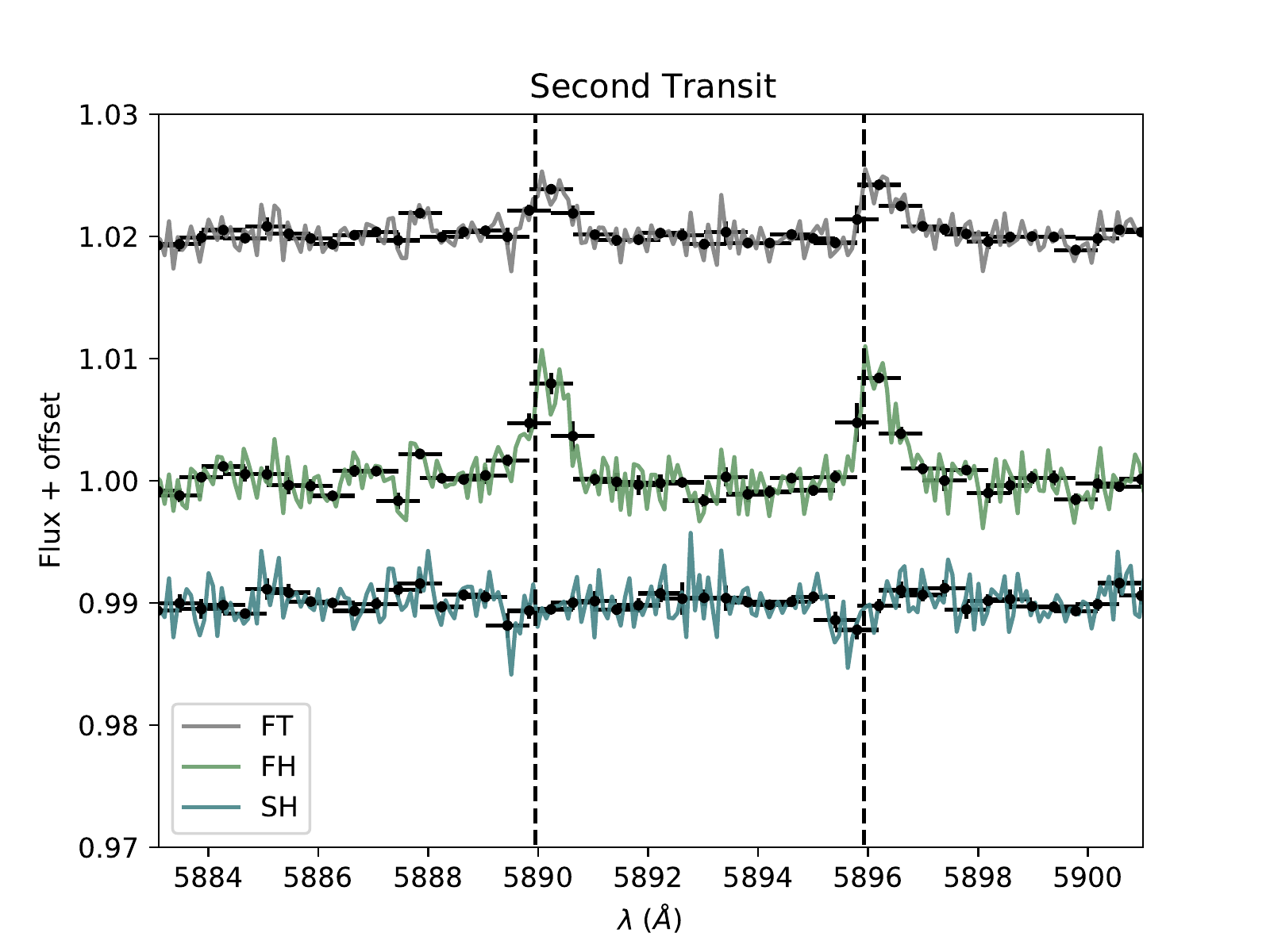}}
    \caption{Top panels: Observed HORuS spectra of KELT-7b around the Na region on T1 (left) and T2 (right) occasions. Black colour stands for the full-transit stacked data (FT); green and blue stand for the planetary spectrum corresponding to the first- and second-half parts of the transit, respectively (labelled as FH and SH). No stellar activity has been removed from the original observations.  The modelled transmission spectrum for KELT-7b is displayed in the top panels a red dashed line (see Sect.~\ref{sec:ulims}). Bottom panels: the synthetic stellar Na lines were modelled as described in the text. They were subjected to the same process analysis as the planetary spectrum extraction. The results are displayed for the full and half transits  (colour coded as in the top panels). The simulated spectra that contain no planetary signal can explain qualitatively the observations.}
    \label{Na_activity}
\end{figure*}

In the light of the observations, the stellar activity and R-M+CLV effects are to a large extent responsible for the $rms$ observed in the planetary spectrum of KELT-7b. They dominate the planetary signal and there are no features related to the atmosphere of the planet (see Figs.~\ref{map_Na}--\ref{map_Ca}). In spite of this, we can calculate  an upper limit on the presence of the lines using the theoretical $rms$ values corresponding to both the stellar activity and R-M+CLV effects. In all, the total observed $rms$ in the transmission spectrum should be the quadratic sum of the R-M+CLV ($rms_{R-M}$) and the stellar activity effects ($rms_{activity}$). Consequently the total observed $rms$ can be explained by the following expression: 
\begin{equation}
\label{eq:rms}
rms_{obs,lim}^2=rms_{R-M}^2+rms_{activity}^2+rms_{corr,lim}^2
\end{equation}
Where $rms_{obs,lim}$  represents the observed upper limit while the corrected limit is represented by  $rms_{corr,lim}$.

 These corrected upper limits are given in the last column of Table~\ref{rms_compare} for each atomic species. We imposed the following upper limits:  0.076 \% on the detection of \ion{Na}{i}~D, 1.4~\% for H$\alpha$, 0.119 \% for \ion{Mg}{i}~b, 0.516 \% for \ion{Ca}{ii}~H\&K, and 0.043 \% for \ion{Li}{i} in the atmosphere of KELT-7b. 

\begin{figure}                                      
\centering                       
    \includegraphics[width=0.5\textwidth]{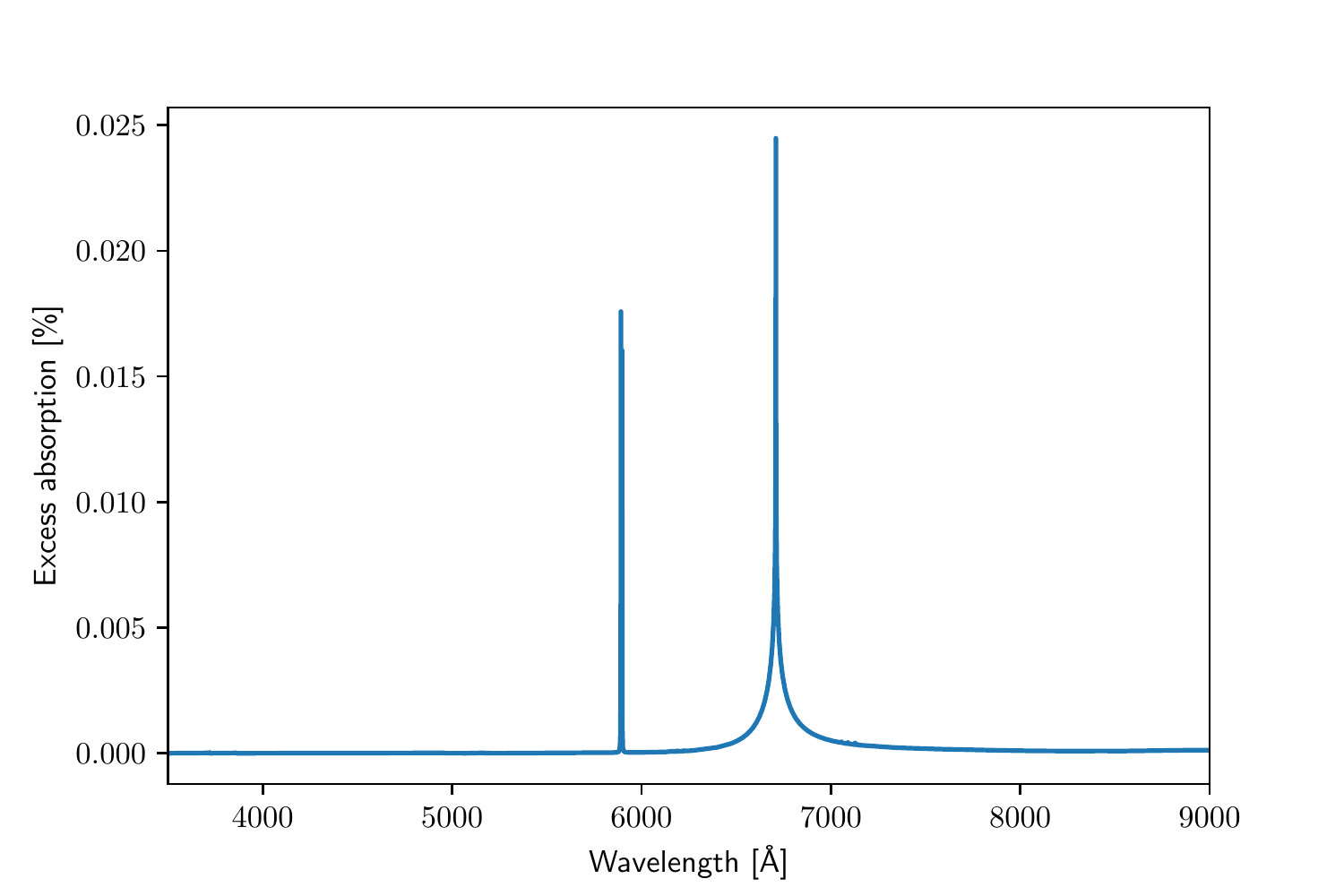}
    \caption{Synthetic model of KELT-7b generated using {\tt petitRADTRANS} (see text for details). The most intense line present in the synthetic spectrum corresponds to the \ion{Li}{i} line at 6707.8\AA{} }
    
    \label{prt_model}
\end{figure}

Finally, we compare these upper limits to the scale-height of the planetary atmosphere. To that aim, using the data listed in Table~\ref{par_planet}, we derived an scale height (H) of 610~$\pm$~115~km that corresponds to an absorption depth of 0.0092~$\pm$0.0018~\%. Thus, the upper limits computed in Table~\ref{rms_compare} correspond to  8.3~H for \ion{Na}{i}~D, 152.2~H for H$\alpha$, 12.9~H for \ion{Mg}{i}~b, 56.1~H for \ion{Ca}{ii}~H\&K, and 4.7~H for \ion{Li}{i} in the atmosphere of the planet. In addition, we can compare these numbers to the predictions of theoretical models. To that aim, we generated a synthetic model of KELT-7b across the HORuS spectral coverage. We employed the {\tt petitRADTRANS}\footnote{\url{https://petitradtrans.readthedocs.io/en/latest/}} \citep{prt} to compute the synthetic model under assuming an isothermal atmosphere. The layer-by-layer composition was computed using the {\tt FastChem} code\footnote{\url{https://github.com/exoclime/FastChem}} \citep{fastchem}. To do the modelling, we assumed the solar abundances of \citep{asp09} scaled with the metallicity of the host star ([Fe/H]~$=$~0.24~$\pm$~0.02~dex, see Table~\ref{par_planet}), although the solar Lithium abundance was assumed to be the meteoric value of 3.3~dex \citep{asp09}. The modelled spectrum only shows two meaningful features in the HORuS wavelength range corresponding to both \ion{Na}{i}~D and \ion{Li}{i} at 6707.8\AA{} (see Fig.~\ref{prt_model}). Only the \ion{Na}{i}~D is covered by the HORuS observations whereas the \ion{Li}{i} line is in an inter-order gap. Further inspection of the KELT-7b synthetic spectrum reveals that the predicted depths of the modelled sodium lines have a depth of approximately 0.02 \% with respect to the continuum. Therefore, we need to improve the noise level at least by a factor of two to be able to probe the Sodium line depths predicted by the {\tt petitRADTRANS} synthetic model.


\section{Conclusions and final remarks}
\label{sec:concfinal}

We analysed two transits of the hot Jupiter KELT-7b using HORuS at the GTC. In this work, we also revised the ephemeris of the KELT-7 system using both photometric and spectroscopic data. We derived new stellar atmospheric parameters (namely, $T_{\rm eff}$, $\log{g}$, [Fe/H], and $\varv \sin{i}$) for KELT-7. Interestingly, it has a $R_{*}$~$=$~1.712~$\pm$~0.037~R$_{\odot}$ that does not correspond to that of an F2~V star in the main sequence. The PARAM web interface provides an stellar age of 1.2~$\pm$~0.7~Gyr that is compatible with a star that is leaving the main sequence towards the subgiant phase. We determined a stellar rotation period of 1.38 $\pm$ 0.05 d using the photometric light curves of three consecutive {\sl TESS} Sectors. Our newly derived planetary ephemeris are compatible with recent determinations in the the literature. In particular, our $T_0$ value implies a shift of  approximately 6 min with respect to the discovery finding reported by \citet{bier15}.\\

In parallel, we constructed a synthetic spectrum in the region 1--1200~\AA{} based on a coronal model mentioned in Sect.~\ref{sec:obs}, following \citet{sf11}. The model-derived fluxes in different EUV bands are 6.2~$\times$~10$^{28}$~erg\,s$^{-1}$ (100--920~\AA{}), and 5.2~$\times$~10$^{28}$~erg\,s$^{-1}$ (100--504~\AA{}), for the first ionisation edges of H and He respectively. We can use the energy-limited formula of mass loss rate \citep[see][and references therein]{sf11} to have an estimation of the photoevaporation in KELT-7~b, resulting in 8.9 $\times$ 10$^{11}$~g\,s$^{-1}$. This value is higher than the mass loss rate calculated for HD~189733 (1.8 $\times$ 10$^{11}$~g\,s$^{-1}$) by the same authors.

We analysed the classical R-M effect employing the RVs corresponding to two HORuS transits alongside a third transit corresponding to the publicly available TRES RVs \citep{bier15}. Our results indicate that KELT-7 has an angle $\lambda$~$=$~0.5\,$^{+2.4}_{-2.9}$~deg  between the sky projections of the star's spin axis and the planet's orbital axis  according to the classical R-M effect. The analysis of the Doppler Shadow results in $\lambda$~$=$~$-$10.55~$\pm$~0.27~deg,  which combined with the planet orbital angle and the near equator-on geometry of the parent star, yields a 3D obliquity of $\psi$ = 12.4 $\pm$ 11.7 deg (or 167.6 deg), thus supporting an aligned planetary system. Thanks to the CCFs calculated using the HORuS observations, we have been able to retrieve the Doppler shadow of KELT-7b and found no trace of an atmospheric signature corresponding to iron/cobalt/nickel, which contrasts with the positive iron detections reported for some UHJ planets. KELT-7b has an equilibrium temperature of 2028~K, which is about 200 K lower than the coolest UHJ planets. This suggests that either neutral iron is not present in gaseous form in the atmosphere of KELT-7b or it remains below the detectability limit of our observations.

We also generated two independent transmission spectra for Kelt-7b. They cover the wavelength regions corresponding to the following atomic lines: H$\alpha$ (6562.78~\AA{}), \ion{Li}{i} (6103.65~\AA{}), \ion{Na}{i} (5889.95 and 5895.92~\AA{}), \ion{Mg}{i} (5167.32, 5172.68~\AA{}), and \ion{Ca}{ii} (3933.66 and 3968.47~\AA{}). The extracted transmission spectra have a significant contribution due to the stellar activity and the R-M effect. These two effects pollute the transmission signal and they only allow us to calculate an upper limit to the presence of H, Li, Na, Mg, and Ca in the planetary spectrum. In addition, we corroborated that even tiny stellar activity changes can create faked absorption features in the extracted planetary transmission spectrum that can mimic those of a planet signature. In our data, and particularly focusing on the Na features, the faked signatures are blue-shifted by a small velocity, a property that is typically ascribed to planetary winds \citep[see, e.g.][]{cas19,ehr20} provided their planetary origin were confirmed. To overcome the difficulty added to the analysis by the stellar activity, observing the planetary transits in two or more occasions is strongly advised. Also recommended is to explore species less affected by stellar activity.

\section*{Acknowledgements}
 The authors acknowledge the HORuS project team for its effort and dedication to the HORuS instrument. We acknowledge financial support from the Agencia Estatal de Investigaci\'on of the Ministerio de Ciencia, Innovaci\'on y Universidades through projects PID2019-109522GB-C51,54, and the Centre of Excellence ``Mar\'ia de Maeztu'' award to Centro de Astrobiolog\'ia (MDM-2017-0737). CAP, JIGH and RR acknowledge financial support from the Spanish Ministry MICINN projects AYA2017-86389-P, PID2020-117493GB-I00, and the Spanish Ministry of Science and Innovation under the FEDER Agreement INSIDE-OOCC (ICTS-2019-03-IAC-12). JIGH also acknowledges financial support from the Spanish Ministry of Science and Innovation (MICINN) under the 2013 Ram\'on y Cajal program RYC-2013-14875. This research has made use of the NASA Exoplanet Archive, which is operated by the California Institute of Technology, under contract with the National Aeronautics and Space Administration under the Exoplanet Exploration Program.  This work has made use of data from the European Space Agency (ESA) mission {\it Gaia} (\url{https://www.cosmos.esa.int/gaia}), processed by the {\it Gaia} Data Processing and Analysis Consortium (DPAC,\url{https://www.cosmos.esa.int/web/gaia/dpac/consortium}). Funding for the DPAC has been provided by national institutions, in particular the institution participating in the {\it Gaia} Multilateral Agreement. This research made use of the Vienna Atomic Line Database operated at Uppsala University, the Institute of Astronomy RAS in Moscow,and the University of Vienna. This paper includes data collected with the TESS mission, obtained from the MAST data archive at the Space Telescope Science Institute (STScI). Funding for the TESS mission is provided by the NASA Explorer Program. STScI is operated by the Association of Universities for Research in Astronomy, Inc., under NASA contract NAS 5–26555. This research made use of Astropy,\footnote{\url{http://www.astropy.org}} a community-developed core Python package for Astronomy \citep{astropy:2013, astropy:2018}. Based on observations made with the Gran Telescopio Canarias (GTC), installed at the Spanish Observatorio del Roque de los Muchachos of the Instituto de Astrofísica de Canarias on the island of La Palma. This research has made use of the SIMBAD database, operated at CDS, Strasbourg, France. This research has made use of the NASA Exoplanet Archive, which is operated by the California Institute of Technology, under contract with the National Aeronautics and Space Administration under the Exoplanet Exploration Program. This publication makes use of VOSA, developed under the Spanish Virtual Observatory project supported by the Spanish MINECO through grant AyA2017-84089.
VOSA has been partially updated by using funding from the European Union's Horizon 2020 Research and Innovation Programme, under Grant Agreement nº 776403 (EXOPLANETS-A).
\section*{Data Availability}
The data underlying this article will be shared on reasonable request to the corresponding author.

\bibliographystyle{mnras}
\bibliography{kelt7} 
\appendix

\section{Appendix}

 
\begin{figure}                           
   \centering                       
   \includegraphics[width=0.5\textwidth]{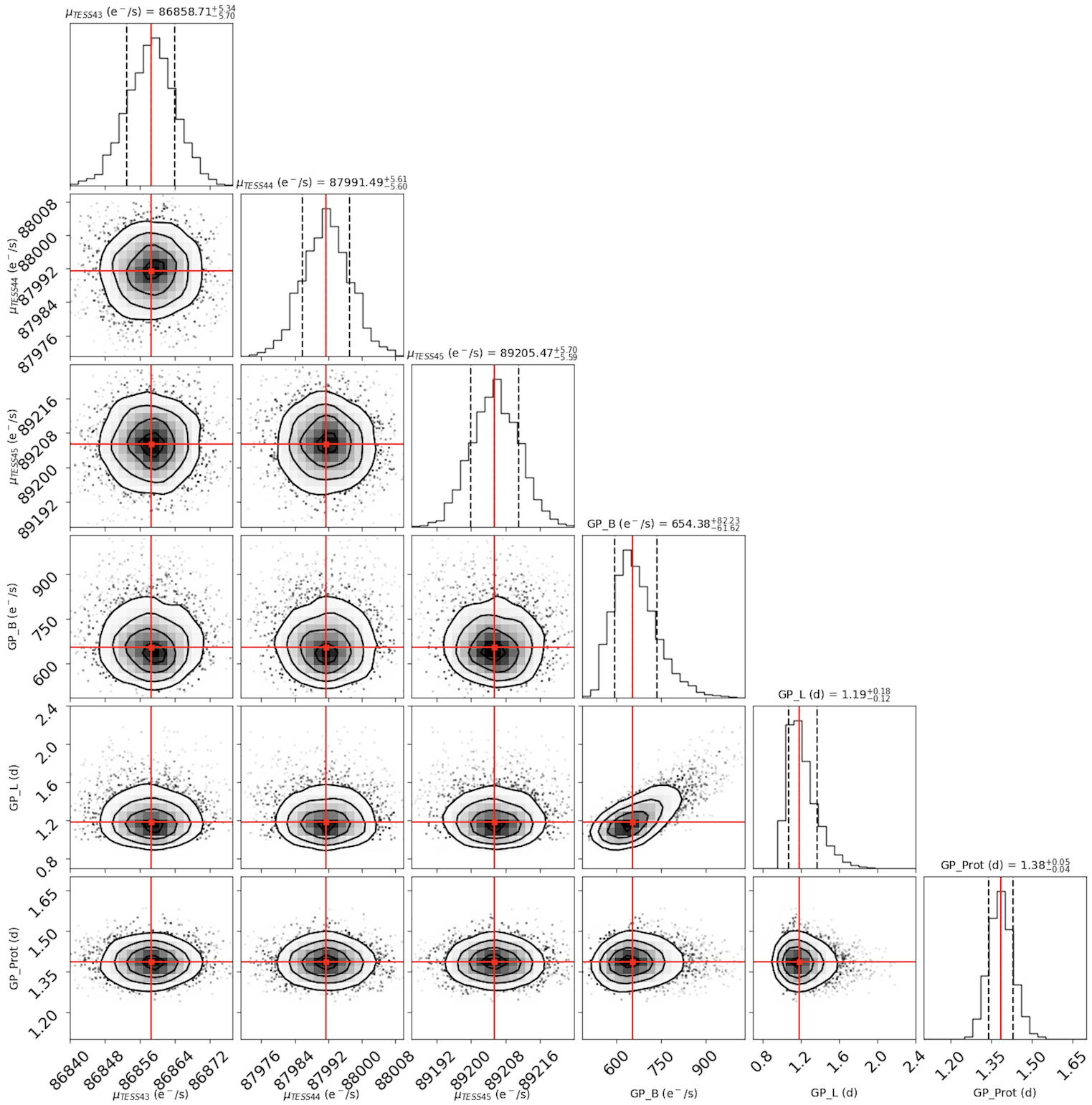}
    \caption{Distribution of the posteriors of the stellar rotation period analysis using the {\sl celerite} quasi-periodic covariance function. The vertical dashed lines indicate the 16, 50, and 84 \% quantiles of the distribution. The mean {\sl TESS} fluxes for each Sector and the kernel parameters, including KELT-7's rotation period, are represented with the $\mu$, GP$_B$, GP$_L$, and GP$_{P_{rot}}$ nomenclature.}
    \label{prot_corner}
\end{figure}
    
      \begin{figure}                           
   \centering                       
   \includegraphics[width=0.5\textwidth]{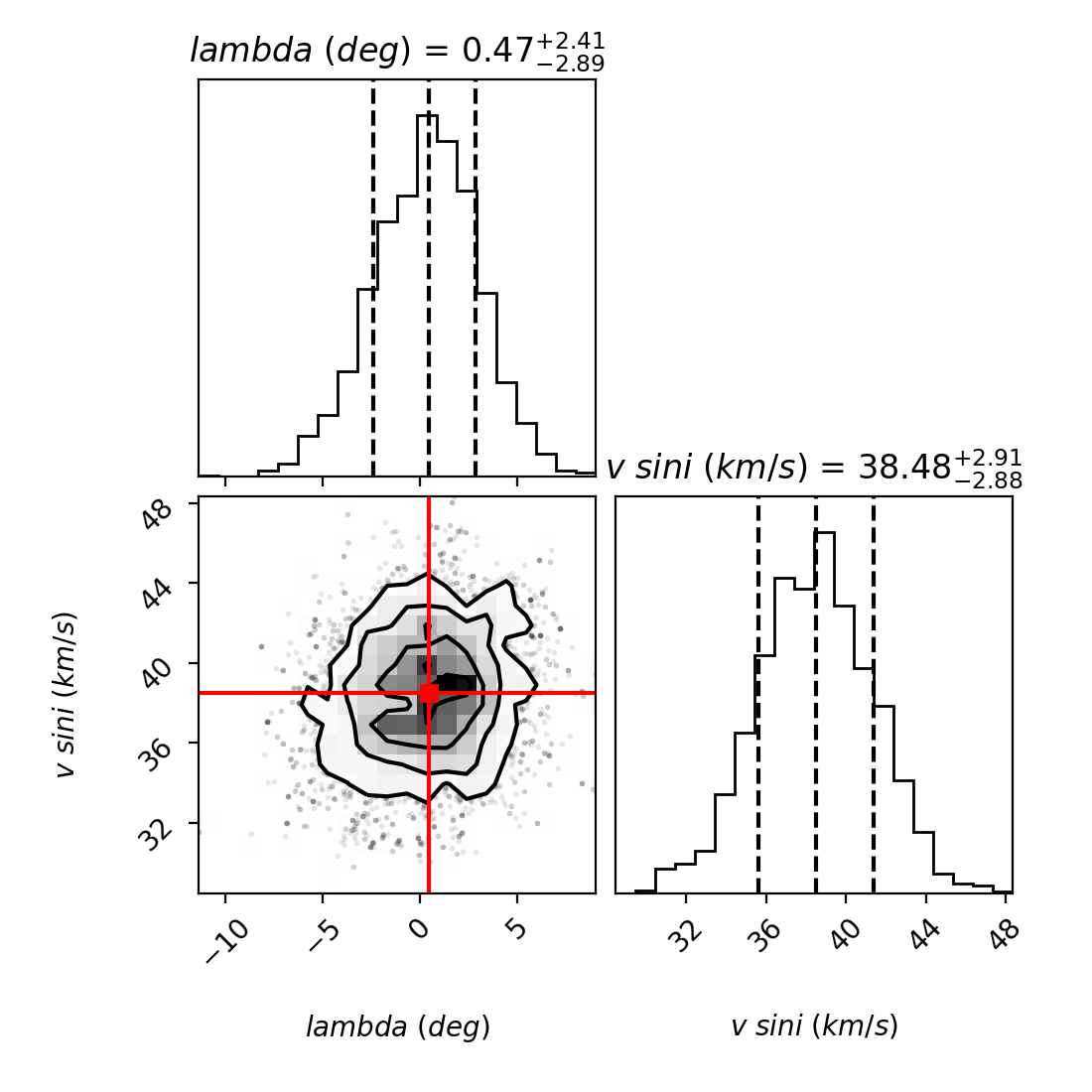}
    \caption{Distribution of the posteriors of the R-M effect analysis using the ARoME code. The vertical dashed lines indicate the 16, 50, and 84 \% quantiles of the distribution.}
    \label{rm_corner}
\end{figure}

\begin{figure*}                                      
\centering                       
   \centerline{\includegraphics[width=0.4\textwidth]{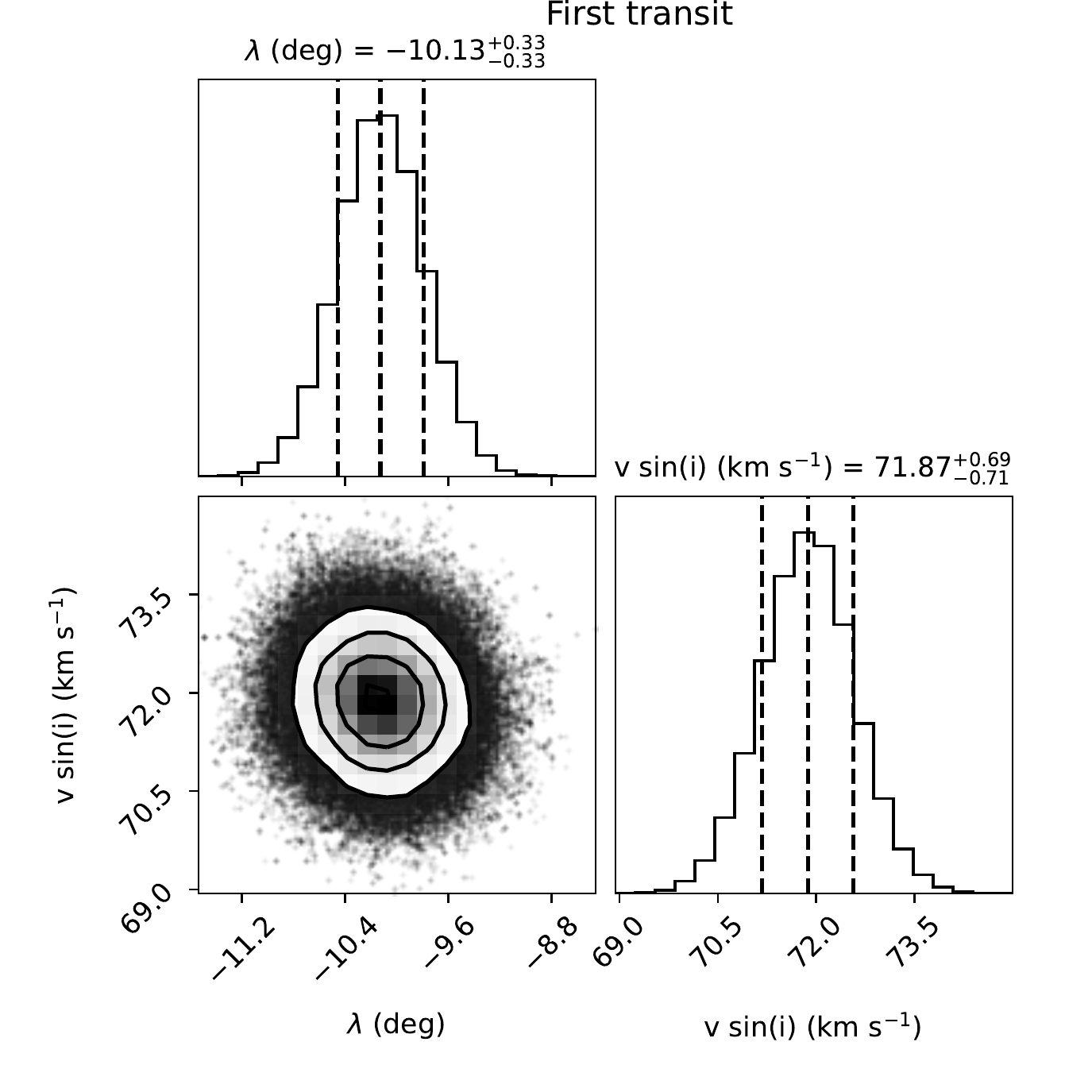}
   \includegraphics[width=0.4\textwidth]{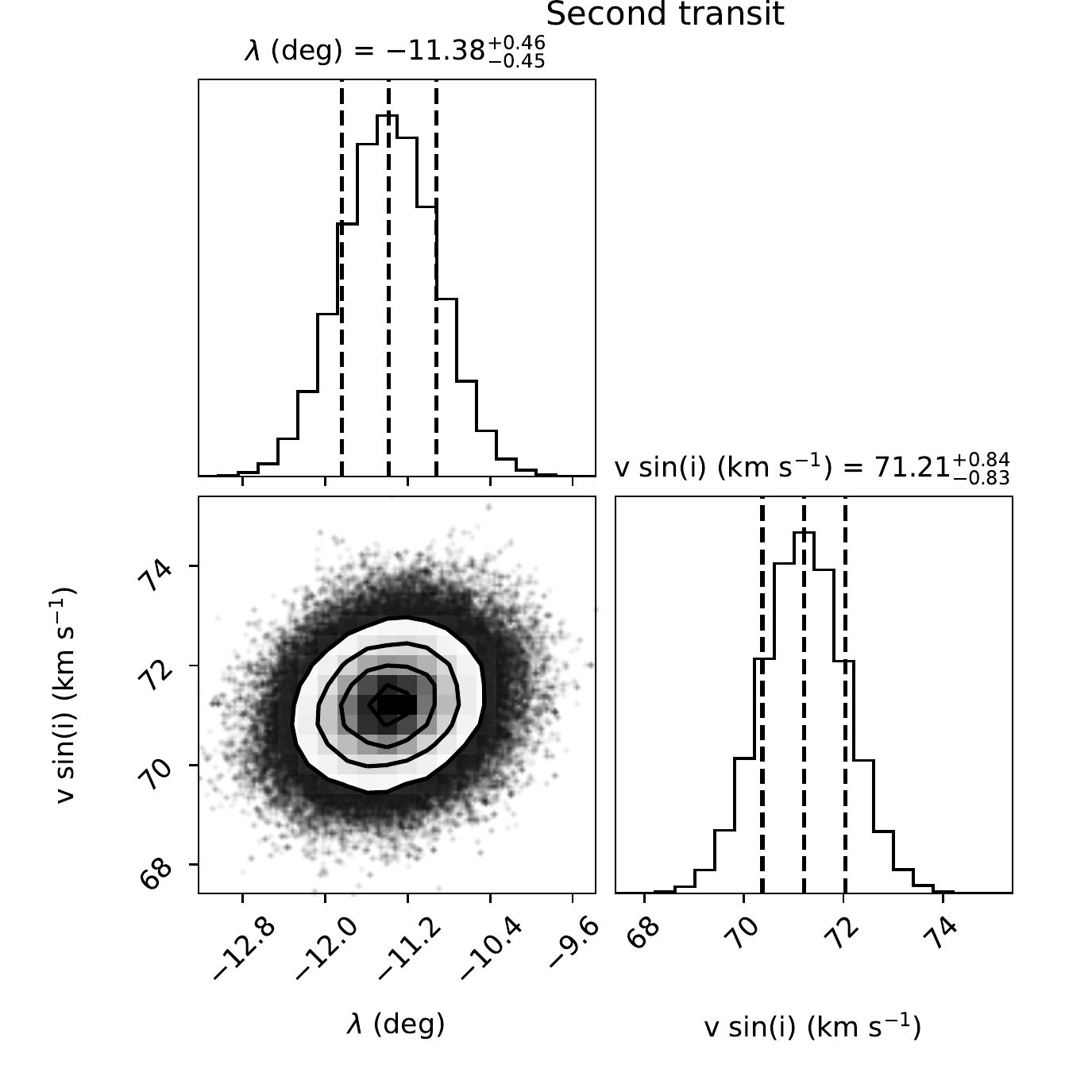}}
    \caption{Distribution of the posteriors corresponding to the Doppler Shadow analysis of KELT-7b  for both transits. The vertical dashed line indicate de 16,50, and 84\% quantiles of the distribution.}
    \label{dopsha_corner}
\end{figure*}

      \begin{figure*}                           
\centering                       
   \centerline{ \includegraphics[width=0.4\textwidth]{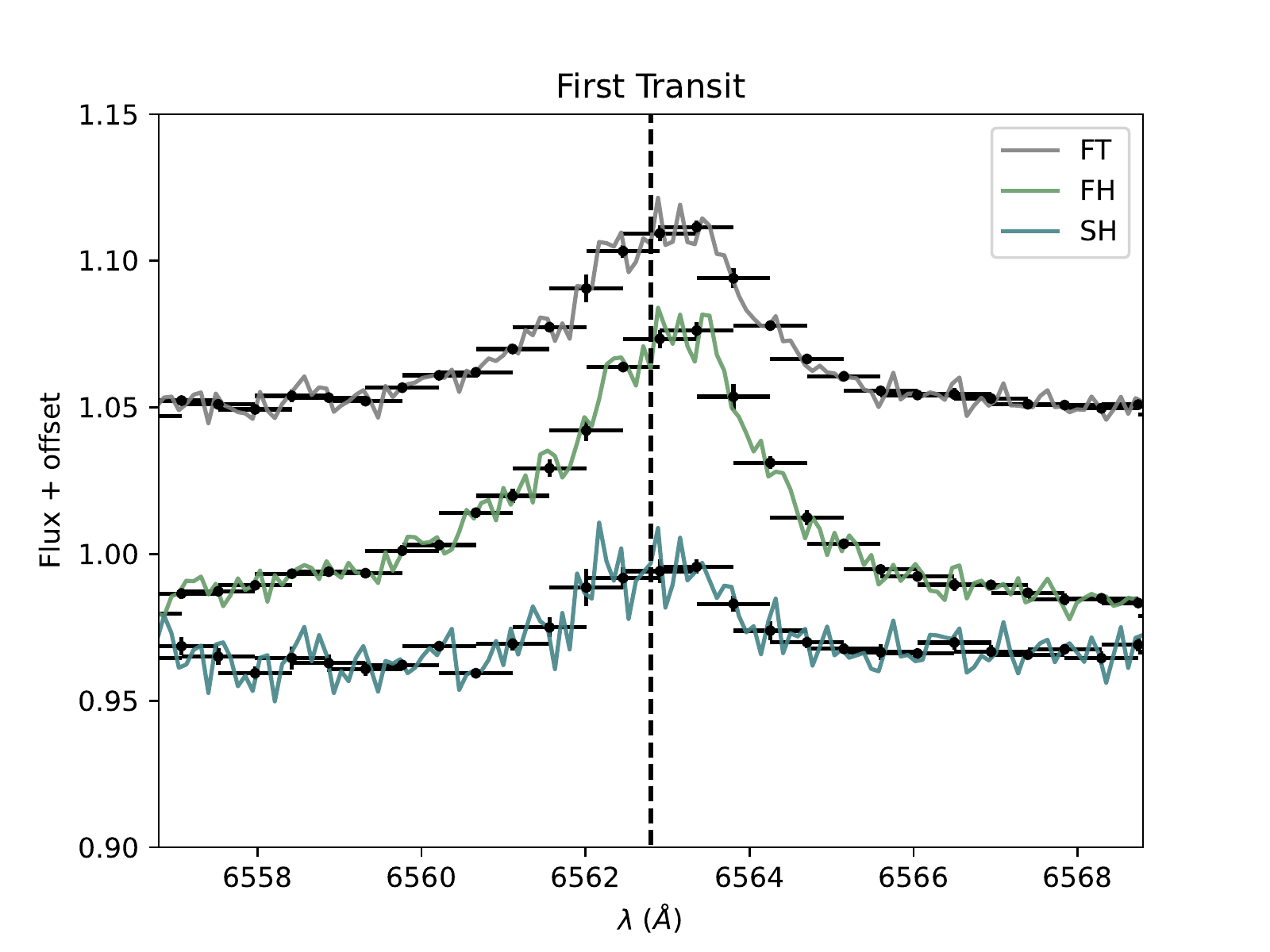}
    \includegraphics[width=0.4\textwidth]{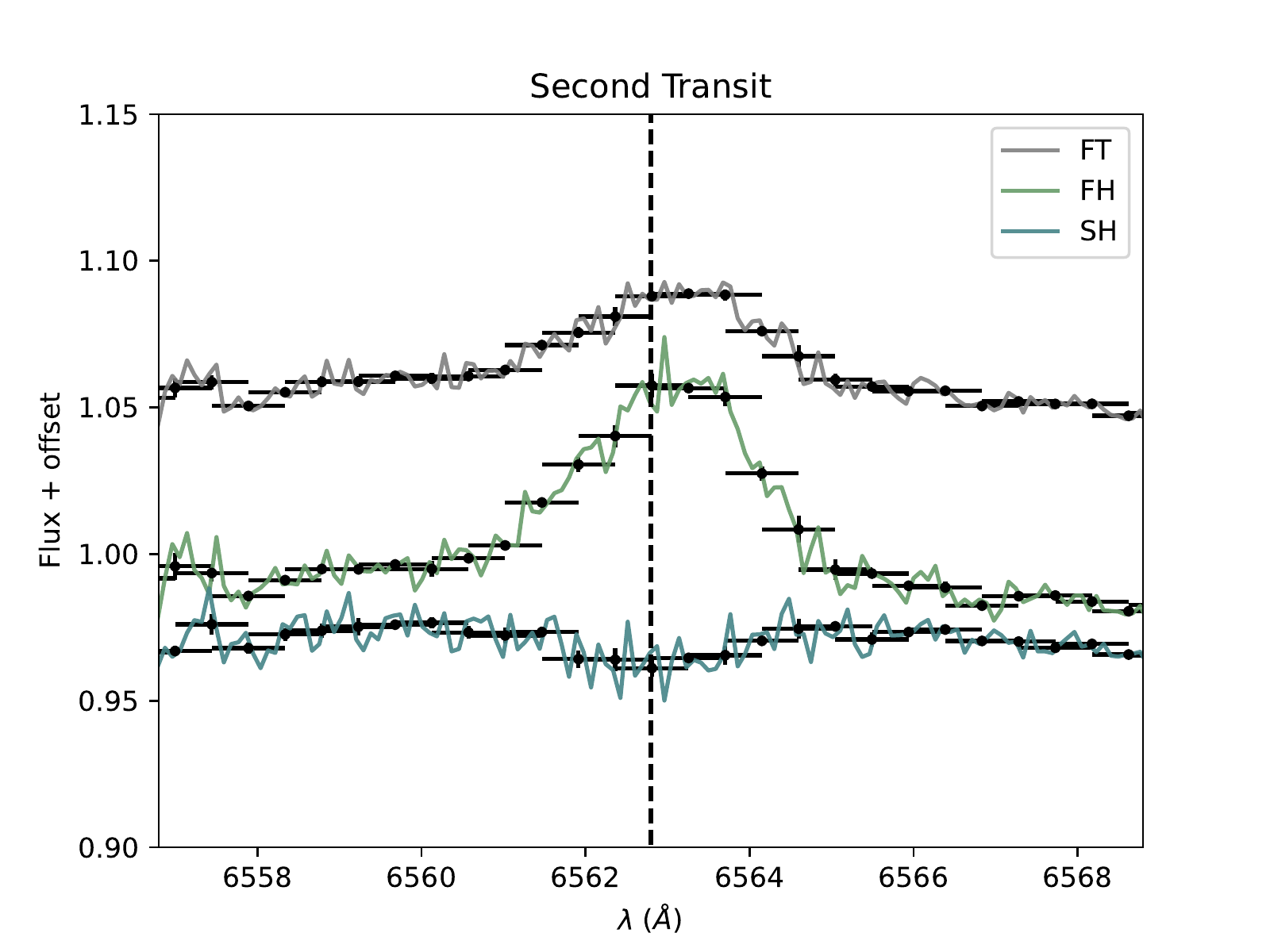}}
       \centerline{ \includegraphics[width=0.4\textwidth]{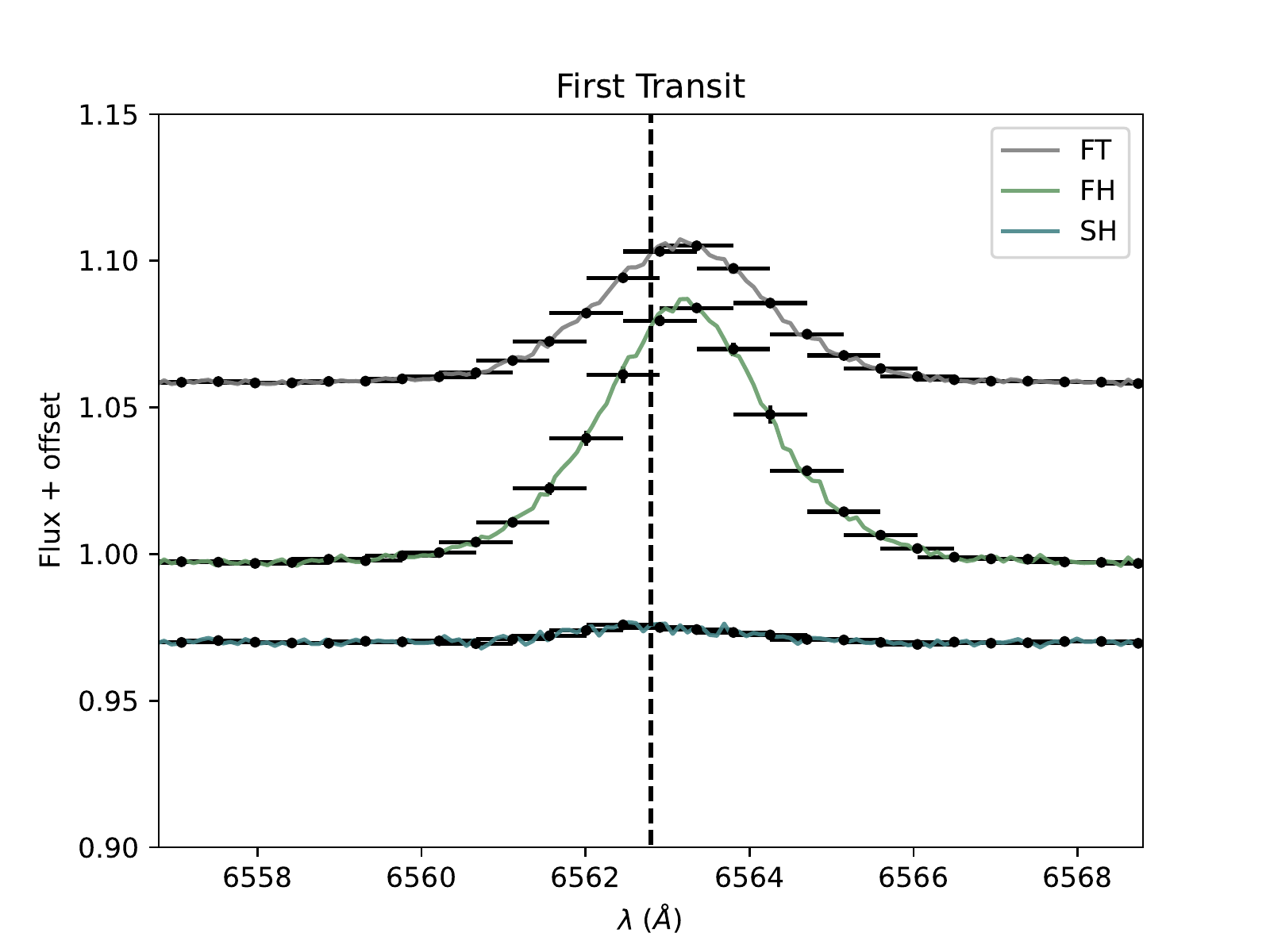}
    \includegraphics[width=0.4\textwidth]{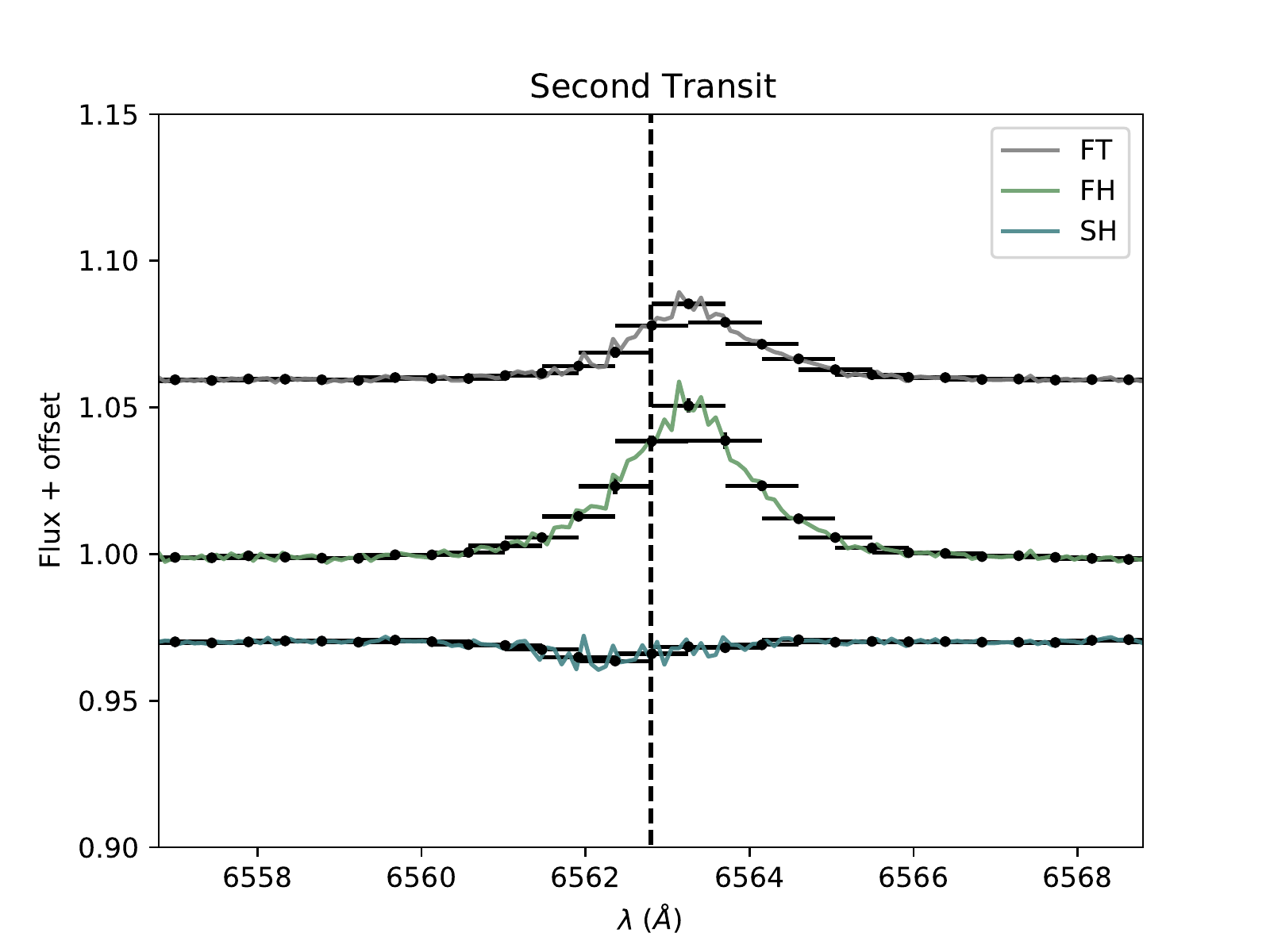}}
    \caption{Same as Fig.~\ref{Na_activity} but for H$\alpha$. Legend is that of Fig.~\ref{Na_activity}.}
    \label{trans_features_Ha}
\end{figure*}

\begin{figure*}                                      
\centering                       
   \centerline{ \includegraphics[width=0.39\textwidth]{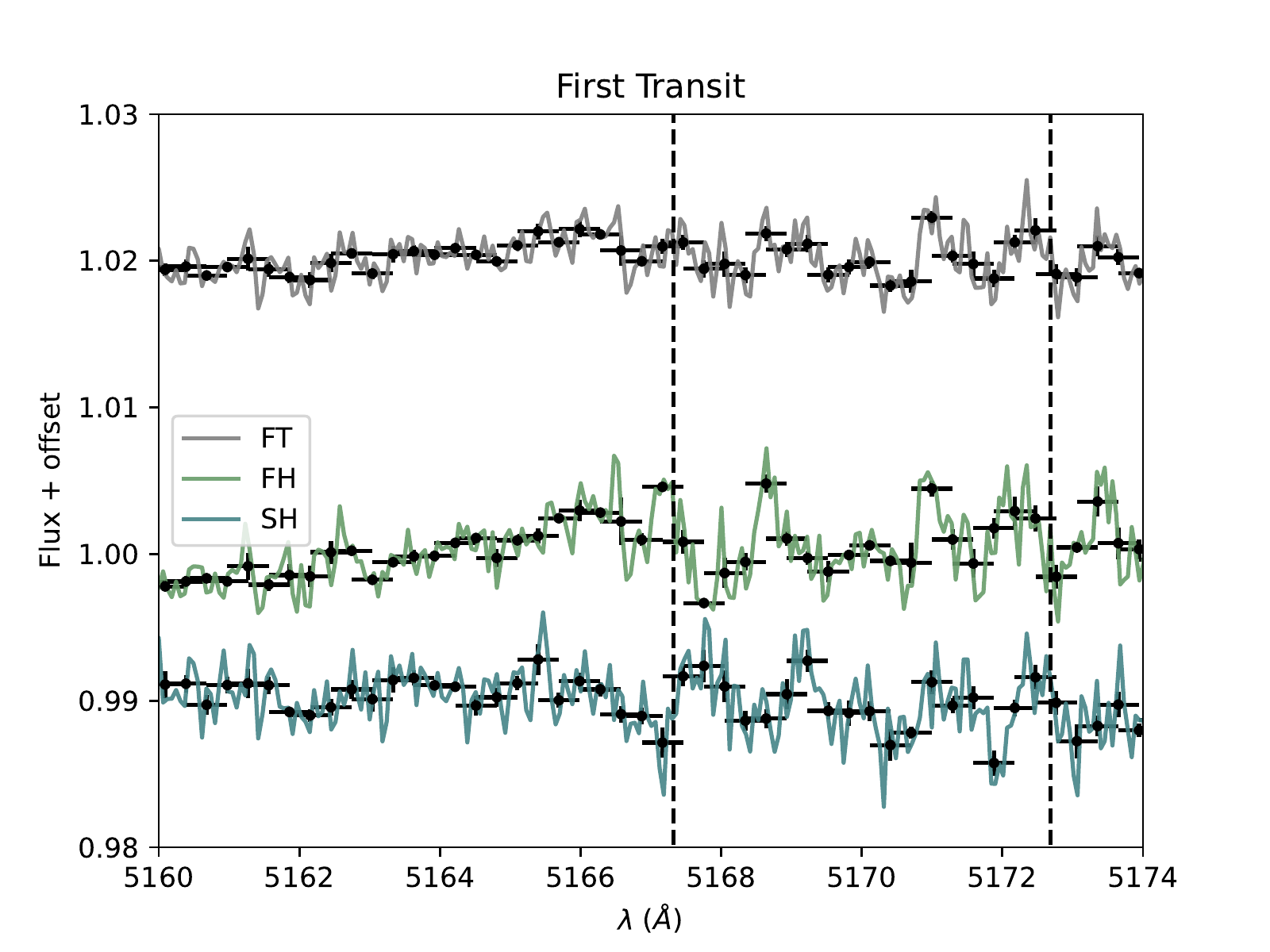}
    \includegraphics[width=0.39\textwidth]{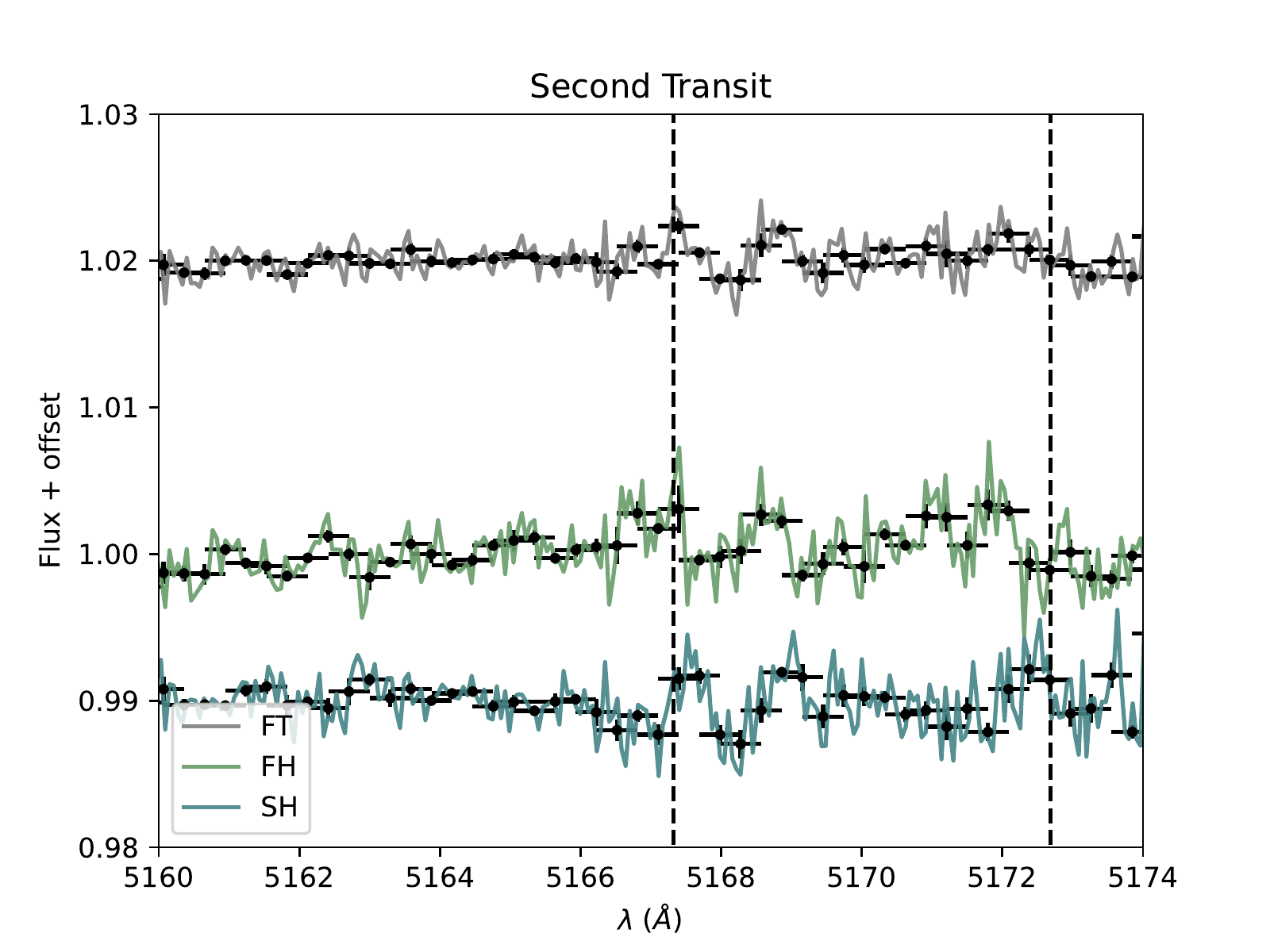}}
       \centerline{ \includegraphics[width=0.38\textwidth]{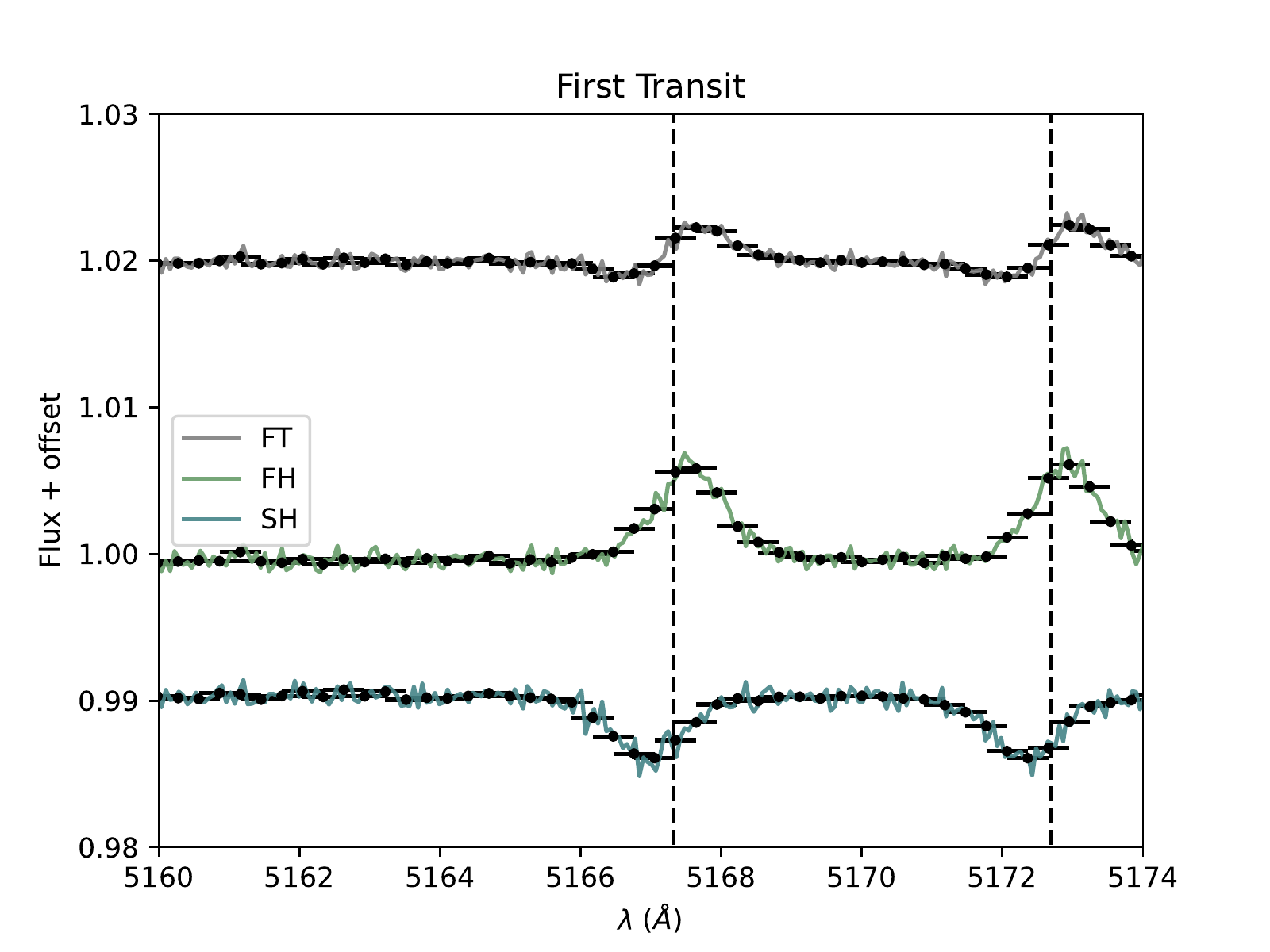}
    \includegraphics[width=0.39\textwidth]{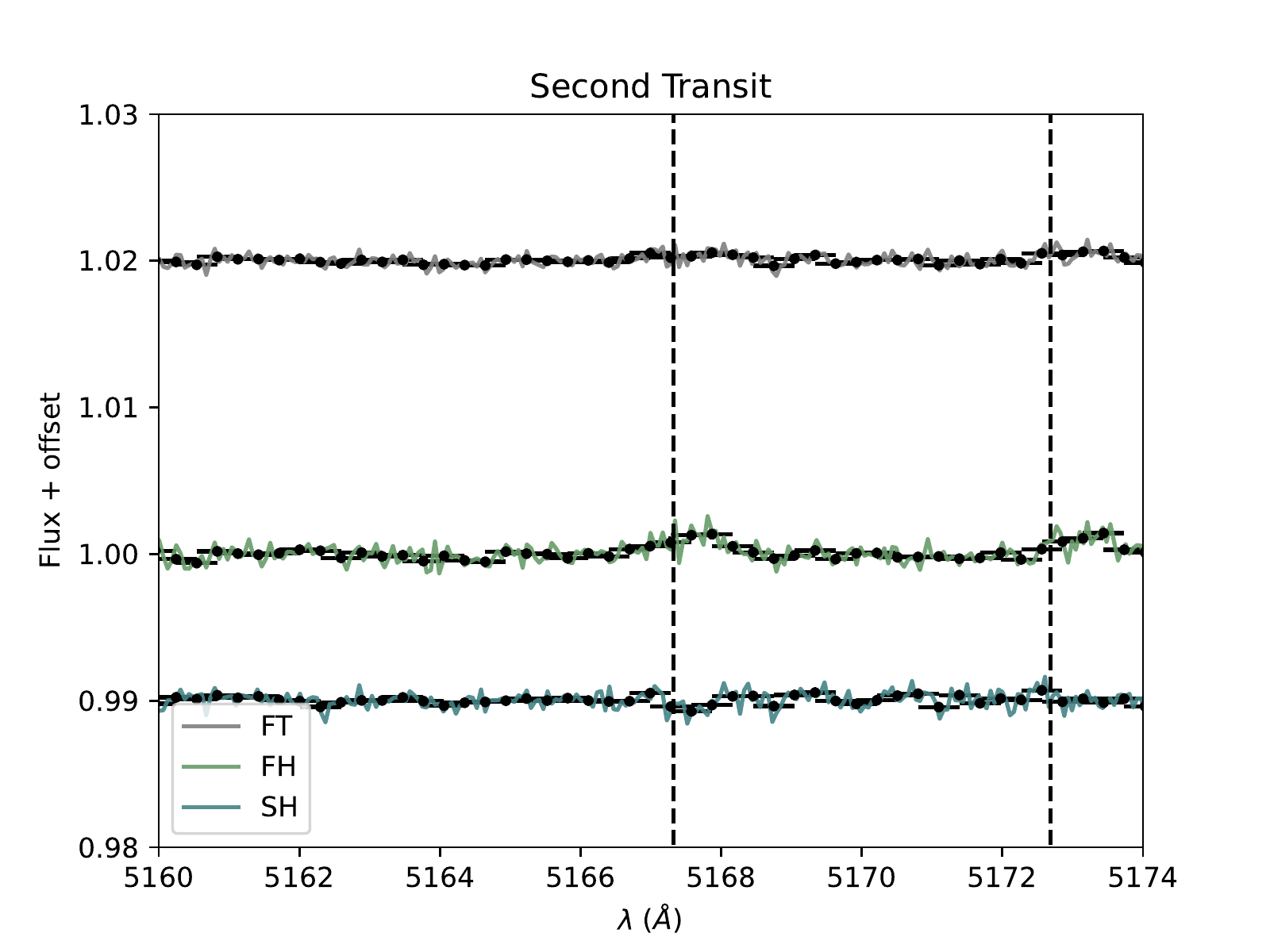}}
    \caption{Same as Fig.~\ref{Na_activity} but for \ion{Mg}{i}. The structures present in the observations are due to the combination of photon noise, the R-M effect, and stellar activity. The simulated spectra only take into account stellar activity and the photon noise.  Legend is that of Fig.~\ref{Na_activity}.  }
    \label{trans_features_Mg}
\end{figure*}   

\begin{figure*}                                      
\centering                       
   \centerline{ \includegraphics[width=0.39\textwidth]{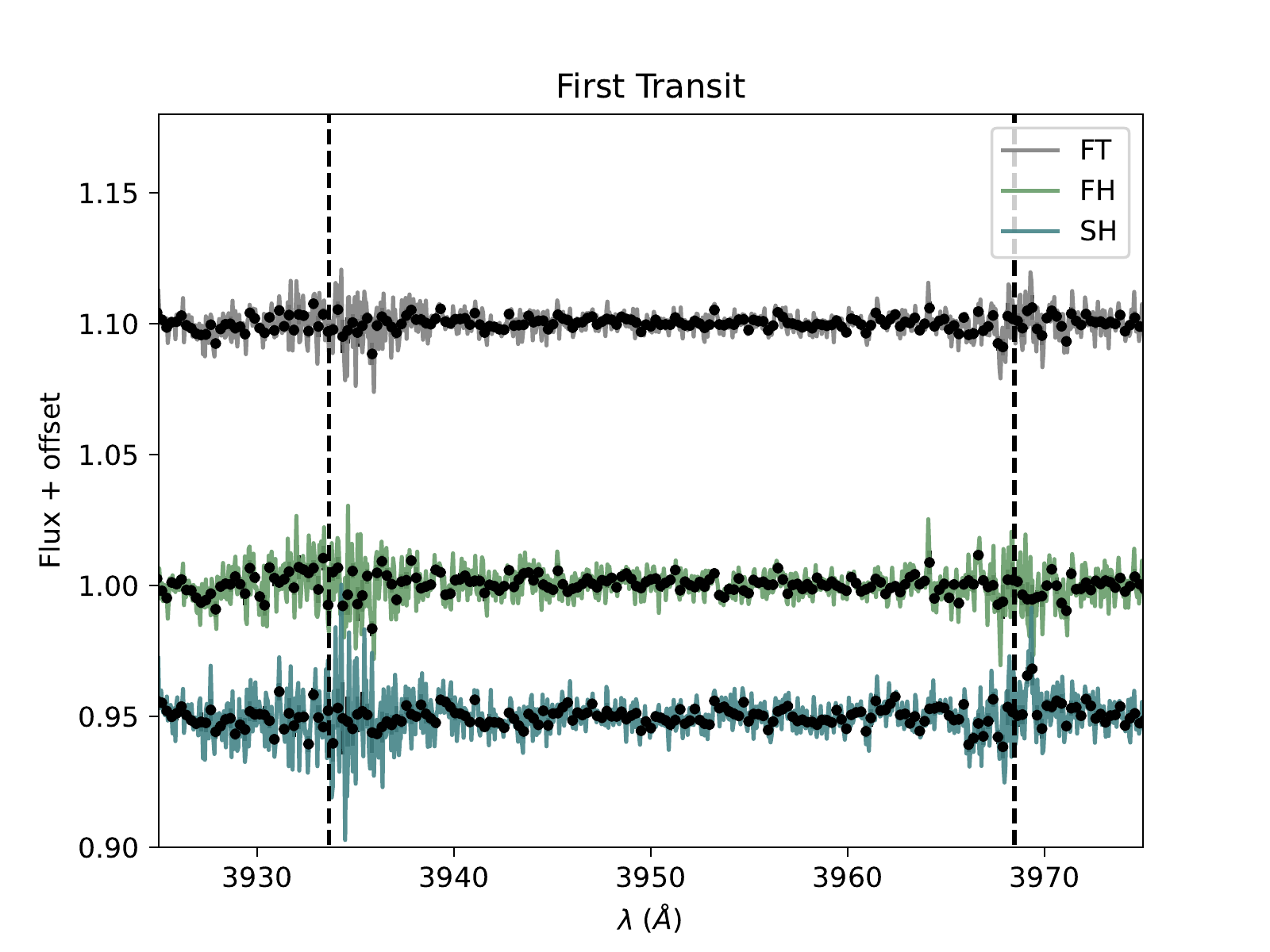}
    \includegraphics[width=0.39\textwidth]{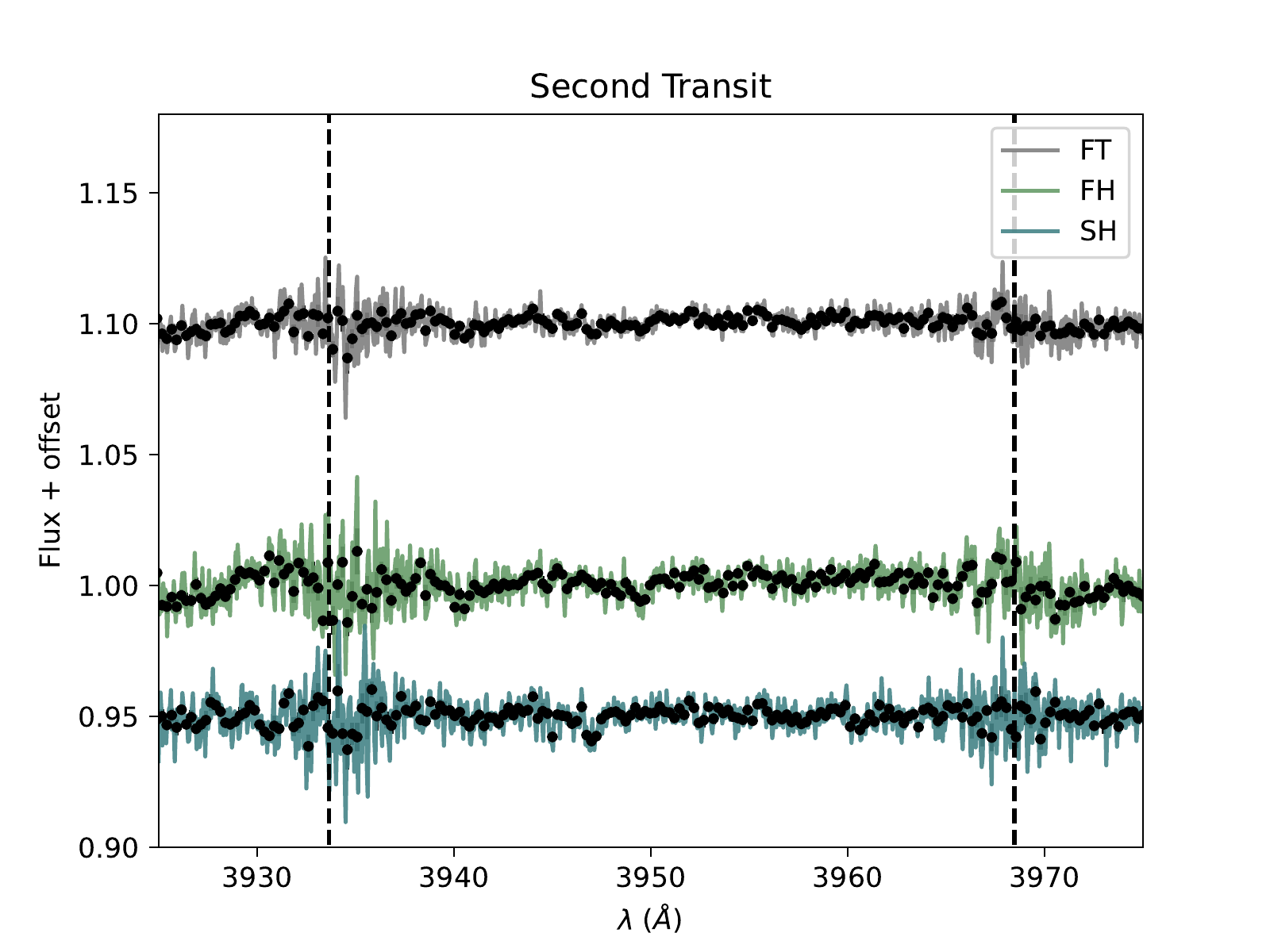}}
       \centerline{ \includegraphics[width=0.39\textwidth]{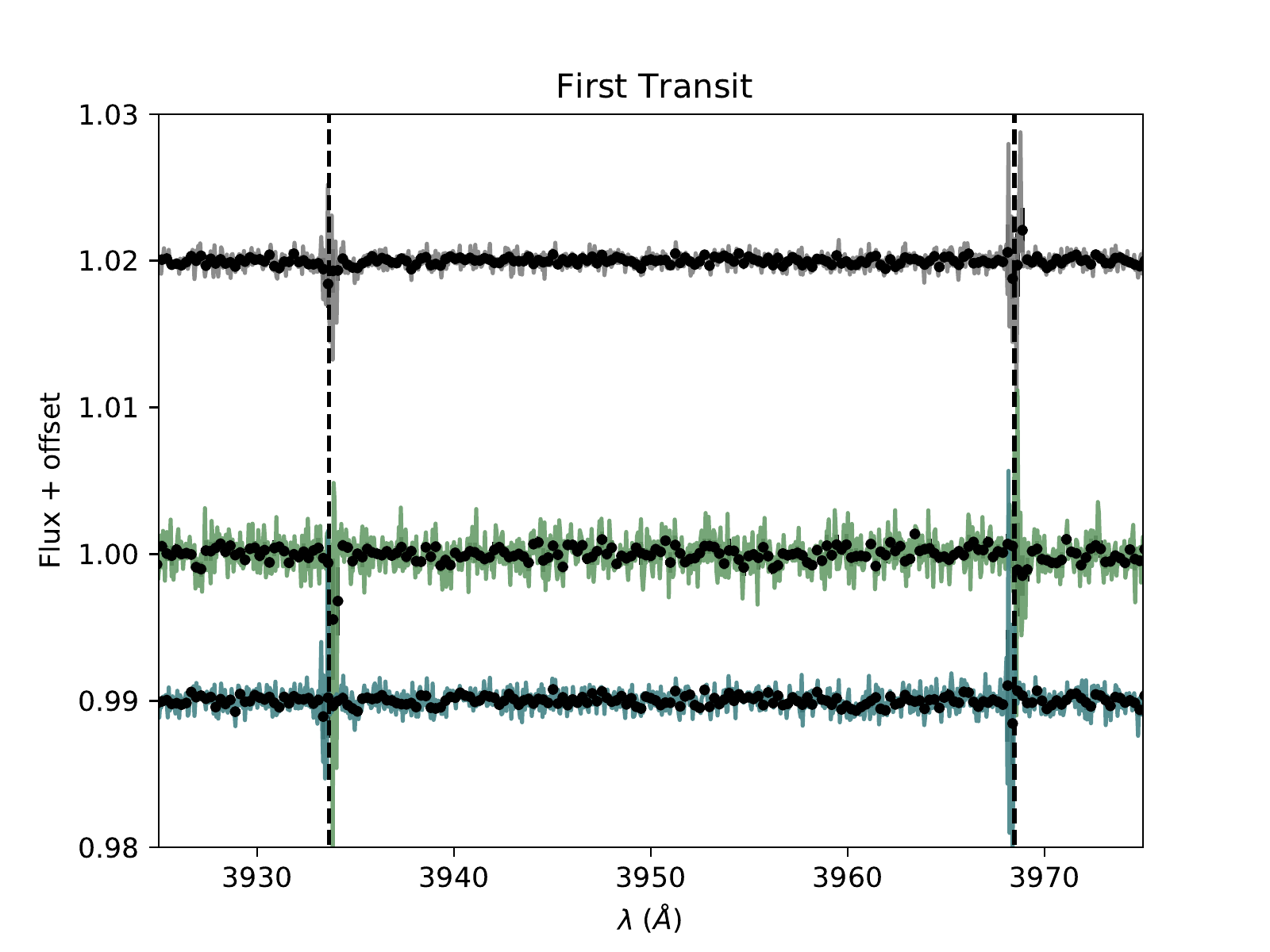}
    \includegraphics[width=0.39\textwidth]{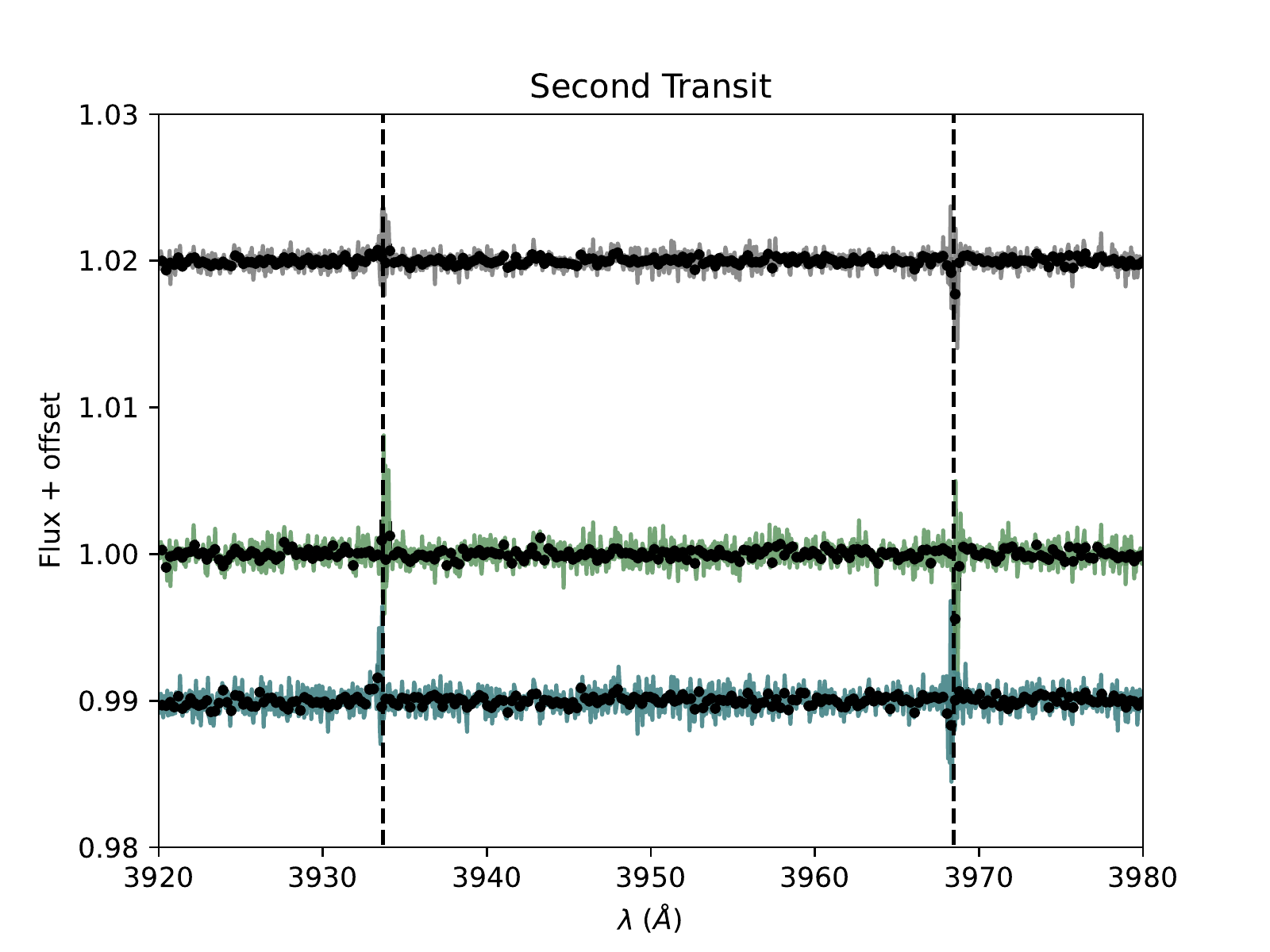}}
    \caption{Same as Fig.~\ref{Na_activity} but for \ion{Ca}{ii}. Legend is that of Fig.~\ref{Na_activity}. }
    \label{trans_features_Ca}
\end{figure*}   

\begin{figure*}
\centering
\includegraphics[scale=0.3]{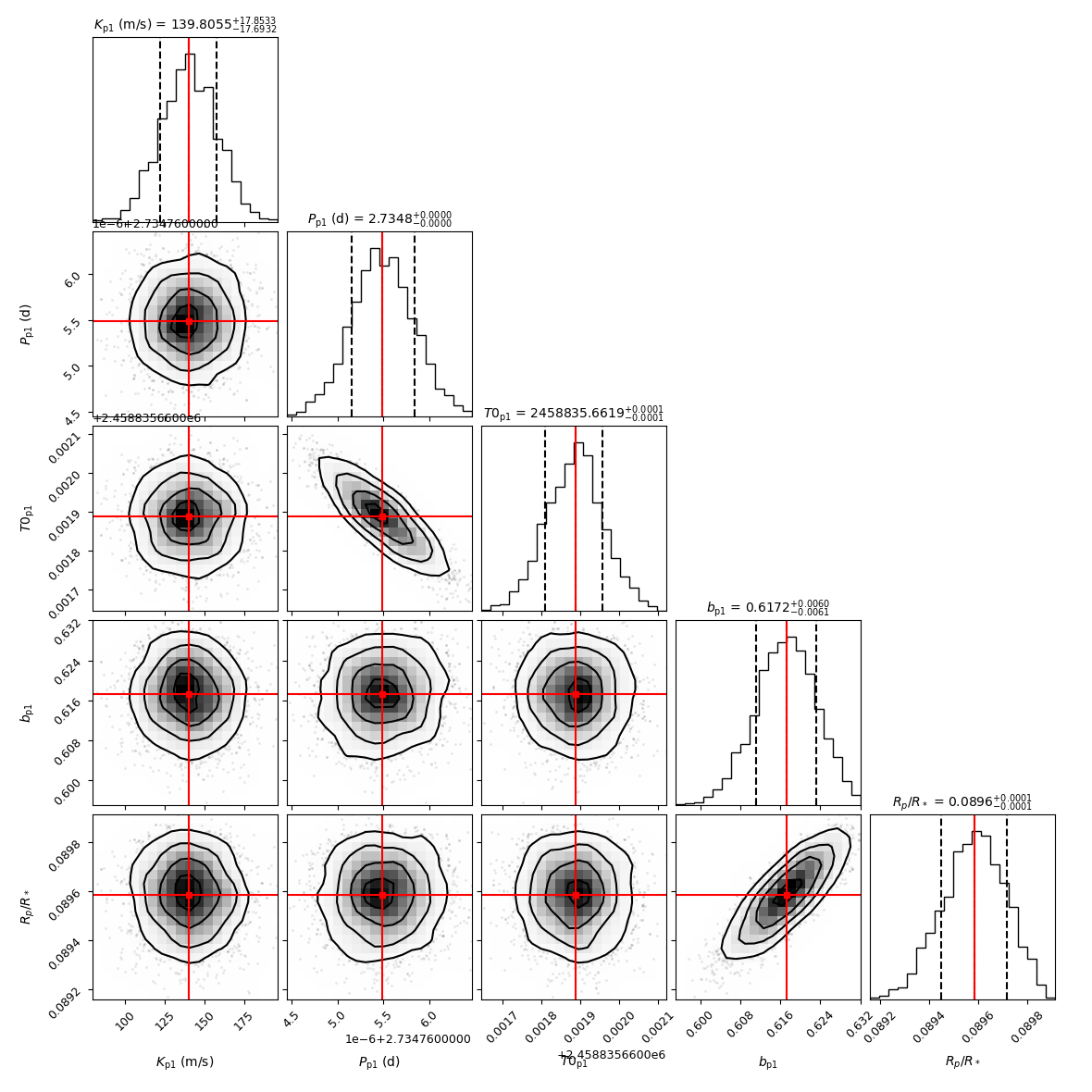}
\caption{Posteriors distributions of the parameters of the KELT-7 light curve and RV joint fit. The vertical dashed lines indicate the 16, 50, and 84\,\% ~quantiles of the fitted parameters; this corresponds to 1$\sigma$ uncertainty. The  red line shows the median values.  We also show the scaled radius ($R_{\rm p}/R_*$) and the impact parameter ($b$) derived from the fit.
}
\label{fig:kelt7_cornerplot}
\end{figure*}

\begin{table*}
\label{tab:obs}
\centering
\footnotesize
\caption{Observational time series for KELT-7. We provide Modified Julian Dates (MJD), airmasses,  radial velocities (RVs), and the activity indices. The full version of this table can be found in the online version of this manuscript.}
\begin{tabular}{lcccccc}
\hline\hline
  MJD  & airmass & RV             &  S-index & Na & Mg & H$\alpha$\\
        &         & [km~s$^{-1}$]  &          &    &    &          \\
\hline
       58805.00274 & 1.191 & 38.07~$\pm$~0.19 &  0.130 & 0.862 & 1.338 & 0.331 \\ 
       58805.00631 & 1.177 & 38.64~$\pm$~0.19 &  0.130 & 0.866 & 1.371 & 0.341 \\ 
       58805.00988 & 1.163 & 38.47~$\pm$~0.19 &  0.126 & 0.868 & 1.372 & 0.334 \\
       58805.01346 & 1.151 & 38.35~$\pm$~0.19 &  0.125 & 0.859 & 1.351 & 0.305 \\
       58805.01703 & 1.139 & 38.24~$\pm$~0.19 &  0.136 & 0.856 & 1.377 & 0.313 \\ 
       ...         &  ...  & ...              &   ...  &  ...  &  ...  &  ... \\
       \hline
       \end{tabular}
       \end{table*}

\begin{table*}
\centering
\footnotesize
\caption{Priors used for the joint LC and RVs fit of KELT-7. The prior labels of $\mathcal{N}$, $\mathcal{U}$, and $\mathcal{L}$ $\mathcal{U}$ represent normal, uniform, and log-uniform distribution, respectively. The error on the density of the star was determined from the stellar mass and radius while the upper limit on the LCs jitter term were set to the $rms$ of the data.}
\label{tab:kelt7_priors_details}
\begin{tabular}{l c c r}

\hline
\hline
\noalign{\smallskip}
Parameter & Prior & Unit & Description \\
\noalign{\smallskip}	
\hline	
\noalign{\smallskip}
\multicolumn{4}{c}{\textit{Stellar parameter}} \\
\noalign{\smallskip}
$\rho_{\star}$ 						& $\mathcal{N}$(426.21, 35.47) & (kg\,m$^{-3}$) & Stellar density\\

\multicolumn{4}{c}{\textit{Photometric parameters}} \\
\noalign{\smallskip}

$\gamma_{\rm TESS}$ 						& $\mathcal{N}$(0, 0.1) & (ppm) & The offset relative flux for the photometric instrument\\
$\gamma_{\rm KELT}$ 						& $\mathcal{N}$(0, 0.1) & (ppm) & The offset relative flux for the photometric instrument\\

$\sigma_{\rm TESS}$ 						& $\mathcal{L} \mathcal{U}$(1e$^{-6}$, 0.004)  & ... & A jitter added in quadrature to the errorbars of instrument\\
$\sigma_{\rm KELT}$ 						& $\mathcal{L} \mathcal{U}$(1e$^{-6}$, 0.03) & ... 	& A jitter added in quadrature to the errorbars of instrument\\

$q1_{\rm TESS}$ 							& $\mathcal{U}$(0, 1)  	& 	...			& Limb-darkening parametrization for photometric instrument\\
$q2_{\rm TESS}$ 							& $\mathcal{U}$(0, 1) &	...		& Limb-darkening parametrization for photometric instrument\\
$D_{\rm TESS}$ 								& 1 (fixed)  					&	...		& The dilution factor for the photometric instrument.\\

$q1_{\rm KELT}$ 							& $\mathcal{U}$(0, 1)  	& 	...			& Limb-darkening parametrization for photometric instrument\\
$q2_{\rm KELT}$ 							& $\mathcal{U}$(0, 1) &	...		& Limb-darkening parametrization for photometric instrument\\
$D_{\rm KELT}$ 								& 1 (fixed)  					&	...		& The dilution factor for the photometric instrument.\\

\multicolumn{4}{c}{\textit{RV parameters}} \\
\noalign{\smallskip}

$\gamma$ 								&  $\mathcal{U}$(-500, 500) & m$\rm s^{-1}$ & RV zero point for TRES\\
$\sigma$		 							& $\mathcal{L} \mathcal{U}$(0.001, 10) & m$\rm s^{-1}$ & Jitter term for TRES \\

\noalign{\smallskip}							    
\multicolumn{4}{c}{\textit{Planet $b$ parameters} }\\	
\noalign{\smallskip}
										    
$P$ 										& $\mathcal{N}$ (2.734, 0.01) 		& d		& Period of planet b \\
$t_0$ (BJD-2,458,000) 						& $\mathcal{N}$ (835.663, 0.1)		& d 		&  Time of periastron passage\\
$e$ 										& 0 (fixed)											& ... 	& Orbital eccentricity of planet b\\
$\omega$ 								& 90 (fixed)										& deg	& Periastron angle of planet b\\
$K$										&  $\mathcal{U}$ (0, 300)  				& 	 m$\rm s^{-1}$     	 & RV semi-amplitude of planet b   \\   
$r_1$									&  $\mathcal{U}$ (0, 1)  					& ... 	     & Parameterization for $p$ and $b$  \\   
$r_2$ 									&  $\mathcal{U}$ (0, 1)  					& ... 	     & Parameterization for $p$ and $b$    \\   
\noalign{\smallskip}	
\hline	
\noalign{\smallskip}
\end{tabular}

\end{table*}


\bsp	
\label{lastpage}
\end{document}